\documentclass[12pt,preprint]{aastex}
\shorttitle{Age and Metallicity Relation of $\omega$ Cen}
\shortauthors{Stanford et al.}
\newcommand{\wcen}{$\omega$~Cen}
\newcommand{\wcent}{$\omega$~Centauri}

\begin{document}

\title{The Age and Metallicity Relation of Omega Centauri}

\author{Laura M. Stanford, Gary S. Da
 Costa and John E. Norris}
\affil{Research School of Astronomy and Astrophysics, Australian
  National University, Weston, ACT, 2611, Australia}
\email{stanford, gdc, jen@mso.anu.edu.au}
\and

\author{Russell D. Cannon} \affil{Anglo-Australian
Observatory, P.O. Box 296, Epping, NSW, 2121, Australia}
\email{rdc@aaoepp.gov.au}

\begin{abstract}
We present a metallicity distribution based on photometry and spectra
for 442 {\wcent} cluster members that lie at the main sequence turnoff
region of the color-magnitude diagram.  This distribution is similar
to that found for the red giant branch.  The distribution shows a
sharp rise to a mean of [Fe/H]~=~--1.7 with a long tail to higher
metallicities.  Ages have then been determined for the stars using
theoretical isochrones enabling the construction of an age-metallicity
diagram.  Interpretation of this diagram is complicated by the
correlation of the errors in the metallicities and ages.
Nevertheless, after extensive Monte-Carlo simulations, we conclude
that our data show that the formation of the cluster took place over
an extended period of time: the most metal-rich stars in our sample
([Fe/H]~$\approx$~--0.6) are younger by 2--4~Gyrs than the most
metal-poor population.  

\end{abstract}


\section{Introduction}
The Galactic globular cluster {\wcent} exhibits unusual properties
compared to other clusters.  The first indication that it was atypical
was in the photometric work of \citet{woo66} and \citet{cs73} where
the large color width of the red giant branch (RGB) was first
established. An internal spread in metallicity was shown to exist from
the spectroscopic work of \citet{fr75} using RR Lyrae stars. A large
range in metallicity from [Fe/H]=--1.8 up to [Fe/H]$\sim$--0.4,
and several discrete populations on the RGB, have been shown to exist
by many studies over the last few decades (\citealt[hereafter
NFM96]{nfm96}; \citealt{sk96,lee99, pan00, rey04, sol05a}).  There
also exist ranges in abundance for all the elements studied in the
cluster \citep{nd95, scl95, smi00}. These studies have primarily
concentrated on the RGB stars as they are brighter than main
sequence~(MS) or subgiant ones.\defcitealias{nfm96}{NFM96}

There is evidence of an age range in {\wcen} from the abundance and
photometric studies.  Specifically, the observed abundance patterns of
different elements show the signatures of a variety of enrichment
processes \citep{le77, nd95, scl95, pan02}.  Contributing sources
include Type~II supernovae which result from high mass stars,
asymptotic giant branch stars (AGB) that lose their material as
stellar winds, and Type Ia supernovae -- formed from older stars via
mass transfer onto a white dwarf. Enrichment of s-process elements is
seen in the RGB stars indicating contributions by low mass
($\sim$1.5--3~solar masses) AGB stars \citep{le77, nd95, scl95,
smi00}. These AGB stars have lifetimes of order 1--3~Gyrs, indicating
that there was an extended period over which enrichment and formation
of the stars occurred in {\wcen}. Results from \citet{pan02} show a
decrease in [Ca/Fe] at higher metallicities in the cluster. This
indicates there are contributions from Type Ia supernovae in the
enrichment processes, and again that it took place over an extended
period.

Using Str\"{o}mgren photometry, \citet{hw00} and
\citet{hr00} examined the metallicity distribution and determined ages
for samples of stars near the turnoff region in {\wcen}. Both
studies concluded that the more metal-rich stars in the cluster were
younger than the metal-poor ones, with an age range of several
gigayears.

Recently, high precision photometry of the cluster has shown perhaps
as many as five discrete RGBs \citep{rey04, sol05a}.  An age range of
$\sim$4 Gyrs was determined using population modeling of the
horizontal branch (HB) by \citet{rey04}. \citet{sol05a} obtained an
upper limit to the range of 6~Gyrs using the RGB bumps which
correspond to the different populations.  The position of the bump in
the RGB luminosity function on the CMD is a function of metallicity
and age (and helium).

\citet{fer04} have shown from their high resolution images that a
distinct metal-rich subgiant branch (SGB) exists (designated SGB-a).
Their conclusion from isochrone fitting the main metal-poor population
and the SGB-a was that they are of the same age, indicating no age
range in the cluster at all.  The distinct SGB-a is also present in
the photometric data of \citet{bed04}.  \citet{bed04} also find two
distinct main sequences but surprisingly the red sequence contains
$\sim$75\% of the stars.  On the RGB the majority of the stars lie
along the blue side of that branch and it has been shown that the
ratio of metal-poor to metal-rich objects is 80:20 \citepalias{nfm96}.
This indicates the bluer sequence is the more metal-rich population,
as has been confirmed spectroscopically by \citet{pio04}. The
separation between the two sequences can be explained by the
populations having significantly different helium abundances
($\Delta$Y$\approx$0.12) \citep{bed04, nor04, lee05}.  The source of
this unusually high helium abundance in the metal-rich population is
not clear.

\citet{hil04} measured abundances of $\sim$400 subgiant and turnoff
stars using medium resolution spectra.  Their abundance distribution
resembles that from the RGB.  Ages were derived for each star using
its metallicity and position on the color-magnitude diagram (CMD)
giving an age-metallicity diagram.  They concluded from this diagram
that a range of about 3~Gyrs exists in the cluster.  Recently
\citet{sol05b} obtained VLT data of $\sim$250 stars on the subgiant
branches in {\wcen}.  They found an age range of no more than 2~Gyrs
fitting isochrones to the populations in the cluster.

It has been suggested by \citet{fre05} that the stars in the most
metal-rich population are actually located in a clump beyond the bulk
of the cluster.  Photometry for this metal-rich population was fitted
with isochrones with metallicities in the range
--1.1$\leq$[Fe/H]$\leq$--0.8 and using a larger distance modulus and
reddening than is conventionally used for {\wcen}.  The direct
spectroscopic abundance measurements of \citet{sol05b} for stars along
the metal-rich subgiant branch found their average metallicity to be
[Fe/H]$\approx$--0.6, casting doubt on the result of \citet{fre05}.

In order to more accurately define the age range in the cluster, we
have observed a sample of MS and turnoff (MSTO) stars. This sample
will enable several new insights into the cluster since we have
studied the MS stars spectroscopically.  We have used these data to
compare the abundance patterns and abnormalities found at the MSTO
with those found for the RGB stars.  The metallicity distribution on
the MS has also been compared to that found on the RGB. A comparison
of our MSTO sample with stars on the RGB can also show whether the
enrichment of s-process and CNO elements is due to surface
contamination, which would be obliterated by the growing convective
envelope as the stars move on to the giant branch, or if the
enrichment is uniform throughout the stars.

The major goal of the present work, however, is to look at the turnoff
region of the CMD along with the metallicities for the members and
determine a more accurate age range for the cluster. Most previous
work at the turnoff region has utilized only photometric data. A
spectroscopic approach coupled with photometry may prove to give a
more accurate age range for the cluster and show any age-metallicity
relation that exists.  In \S2 we describe the observations and
reduction techniques. \S3 outlines the derivation of metallicities for
the sample.  The discussion on the calculation of ages is described in
\S4, and \S5 summarizes the results and comparisons with previous
investigations.  Preliminary accounts of these results have appeared
in \citet{sta04}.

\section{Observations and Reduction} \label{OR_sect}

Photometry for the cluster was obtained with the 1m
telescope/Tektronix CCD combination at Siding Spring Observatory, in
the $V$ and $B$ bands. Ten fields with centers approximately
20~arcminutes from the cluster center were observed. Each field was
20$\times$20~arcminutes in area. Exposure times were 500 seconds for
the $V$ band and 900 seconds for the $B$ band.  Typical seeing ranged
between 1.8--2.2 arcseconds.  The photometry was carried out using
aperture photometry and the final sample only contained uncrowded
stars (i.e. there were no neighbors within 5~arcseconds).  The fields
overlapped slightly in order to calibrate the frames in position and
photometry. The photometry calibration used magnitudes from lists in
the $B$ and $V$ bands from \citet{cs73}, \citet{can81}, and
\citet{cs81} for objects that were in common.  The photometric zero
point uncertainties are of order 0.02 in both bands, and all errors
come from photon statistics.

Preliminary positions were based on an early version of the US Naval
Observatory catalog.  These were then used to match stars from the
SuperCosmos scan of a UK Schmidt plate centered on {\wcen} and
positions of all stars were found.  The final catalog positions from
the SuperCosmos scan have an accuracy of $\sim$0.2~arcseconds.

\begin{figure}
\begin{center}
\includegraphics[width=8.2cm,angle=0]{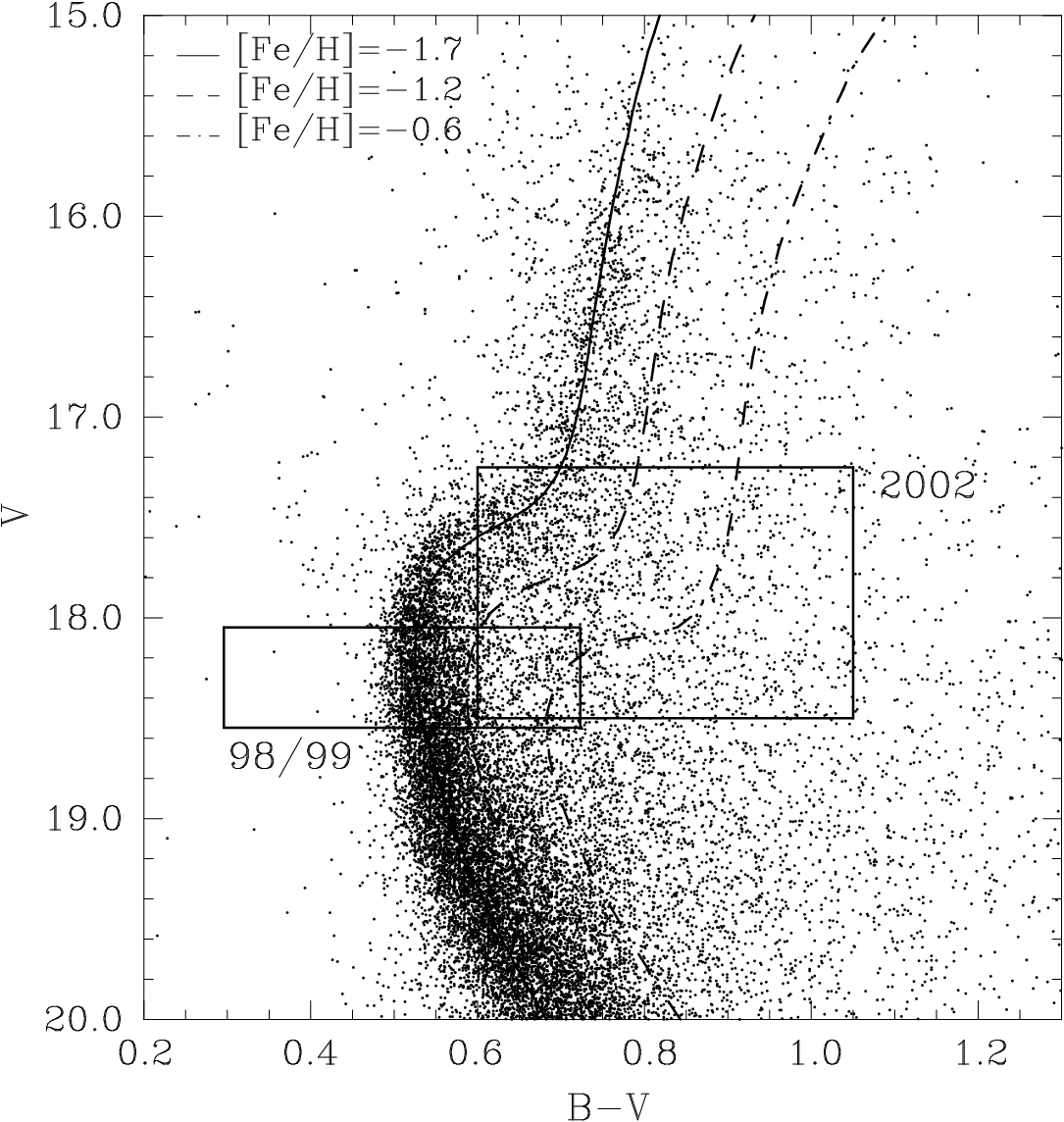}
\figcaption[f1.ps]{ Color magnitude diagram for $\omega$
Cen.  The isochrones \citep{yi01} are for (left to right)
[Fe/H]=--1.7, --1.2, and~--0.6.  Each isochrone has an age of
13.5~Gyrs.  The two boxes represent the areas that were used to find
potential candidates from which to obtain spectra of cluster members.
\label{cmd1} }
\end{center}
\end{figure}

 From these data a CMD, shown in Figure~\ref{cmd1}, was constructed
for objects within an annulus 15--25 arcminutes from the cluster
center. As there is no membership information, Figure~\ref{cmd1}
contains objects that belong to both {\wcen} and the field. The
Yonsei-Yale (Y$^2$) isochrones \citep{yi01} were plotted along with
the data and have metallicities [Fe/H]=--1.7,~--1.2, and~--0.6; all
have an age of 13.5~Gyrs.  Abundance studies have determined that the
different stellar populations in {\wcen} show $\alpha$-enhancement
\citep{nd95, scl95, smi00, pan02, ori03}.  For metallicities
[Fe/H]=--1.7 and~--1.2, [$\alpha$/Fe] was taken to be 0.3, and for
[Fe/H]=--0.6, [$\alpha$/Fe]~=~0.18 was used. In Figure~\ref{cmd1} a
reddening $E(B-V)$=0.11 \citep{lub02} and distance modulus
(m--M)$_{V}$=14.10 were assumed.  This value of the distance modulus
comes from \citet{rey04} and when fitting isochrones best reproduces
the data.  Mean photometric errors as a function of $V$ magnitude from
the aperture photometry are shown in Table~\ref{tbl-1}.

In 1998 an area on the CMD was defined on the upper MS
(18.05${\leq}V{\leq}$18.55; $0.30{\leq}B-V{\leq}0.72$) with a view to
determining the metallicity range.  Stars in this region were observed
using the Two degree Field Multiobject spectrograph (2dF) on the
Anglo-Australian Telescope \citep{lew02}. This spectrograph has the
capability of simultaneously observing up to 400 objects using a fibre
fed system.  In 2002 a second region was defined at the turnoff
(17.25${\leq}V{\leq}$18.5; 0.6${\leq}B-V{\leq}$1.1) to look at the
most metal-rich stars in the cluster and to determine the age range in
the cluster, since the age degeneracy of the isochrones for a given
metallicity can best be broken at the turnoff region. The first sample
was observed in May~1998 and April~1999 (hereafter 98/99 sample), and
the second in March~2002 (hereafter 2002 sample). Figure~\ref{cmd1}
shows the two boxes from which candidates were selected.

Although 2dF is able to observe 400 objects at once, for our sample
the maximum number of objects we were able to observe per
configuration was about 280.  This was due to the compact nature of
our fields relative to the large field-of-view of the instrument, as
well as the limit on fibre-to-fibre spacing.  1200~line/mm gratings
were employed and the spectra obtained covered the useful wavelength
range $\lambda\lambda$3800--4600\AA, with a scale of 1.1{\AA}/pix.
They have a resolution of $\sim$2.4{\AA} {\sc fwhm}. 

As the number of probable members in the 1998 sample was high these
objects were observed in one configuration for several hours.  For the
1999 observations, the small number of non-members found in the
previous run were removed and other candidates added to the
configuration along with the confirmed members.  This single
configuration was again observed for several hours.

A slightly more complicated approach was taken for the 2002
observations.  This sample extended to much redder colors to ensure
than any high metallicity cluster members were included, but
consequently had higher field star contamination. In order to
completely observe our 2002 sample of 900 stars, a number of fibre
configurations were needed and each of these contained successively
fewer new stars due to crowding.  The first step in the observing
process was to determine which stars were members of the cluster.
{\wcen} has a large radial velocity of $\sim$232$\pm$0.7~kms$^{-1}$
\citep{din99a} while the field stars have velocities $\sim$0$\pm$50
kms$^{-1}$. This information was used to determine membership of the
cluster.  Each configuration was observed until a signal-to-noise
ratio per pixel (S/N) of about 10 was reached. The exposure time
depended on the weather and seeing, although the average was 1--2 hrs
per field. The observations were carried out in half hour exposures to
facilitate removal of cosmic rays.  These data were then reduced and
wavelength calibrated using the standard 2dF reduction software
available at the telescope.  The reduced fields were coadded and the
individual spectra were then extracted using the Figaro routine
EXTRACT.  Once reduced and extracted the spectra were cross correlated
with a spectrum of a previously confirmed {\wcen} member in the IRAF
package RV using FXCOR to obtain velocities.  These were plotted as a
histogram and membership was classified above a generous velocity
cutoff limit.

\begin{figure}
\begin{center}
\includegraphics[width=8.2cm,angle=0]{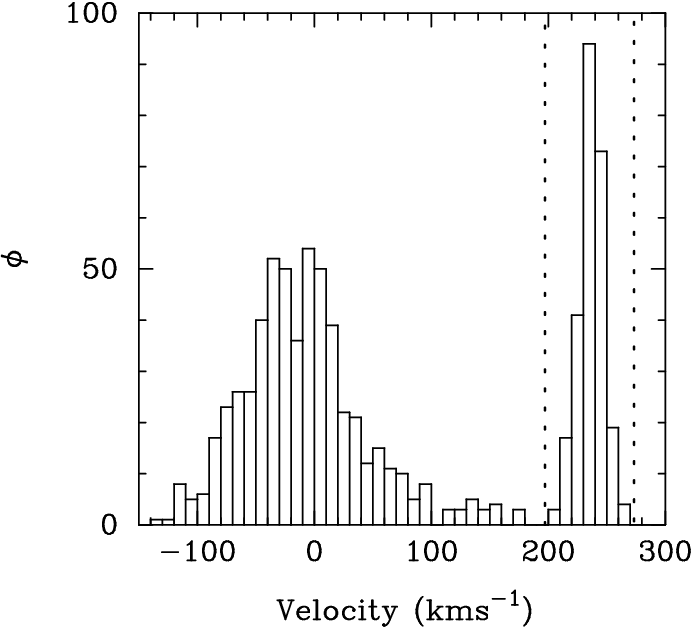}
\figcaption[f2.ps]{Histogram showing the velocities obtain
for all the stars observed in 2002.  The peak at 235~kms$^{-1}$
comprises the {\wcen} members.  The two vertical lines represent the
$\pm3\sigma$ ($\sigma$=13~kms$^{-1}$) cutoff limits applied for
membership classification.
\label{velhis} }
\end{center}
\end{figure}

Once we had observed all stars in our sample, the cluster members were
re-observed with 2dF for $\sim$3--5 hours in order to obtain higher
S/N spectra for more detailed analysis.  The spectra were again
cross-correlated, this time with a synthetic spectrum, to obtain
velocities to confirm membership and obtain more accurate velocities.
Analysis of twilight sky observations taken at the same time showed an
offset in the velocities between the two CCD cameras of
$\sim$11~kms$^{-1}$.  A correction was applied to the spectra to
account for the offset, and the individual spectra then coadded.  The
final heliocentric velocity histogram is shown in Figure~\ref{velhis}.
The narrow peak at 235~kms$^{-1}$ comprises the {\wcen} members while
the broader peak at lower velocities contains the field stars.  The
standard deviation was determined by an iterative $\pm$3$\sigma$
cutoff process (where $\sigma$=13~kms$^{-1}$).  The velocity
dispersion in the outer regions of the cluster is low
($\sim$9~kms$^{-1}$) \citep{mmm97}, and the standard deviation is
driven by both the velocity error measurement and dispersion.  The
mean velocity errors are typically 8~kms$^{-1}$.  The standard
deviation of the field stars was $\sim$50~kms$^{-1}$.

The 98/99 and 2002 samples were observed to completeness levels of
37\% and 94\% respectively, where the completeness level is defined as
the ratio of number of objects observed to the number that had the
potential to be observed. To be classed as ``observable" each star
underwent a visual inspection on the CCD images to ensure there were
no contaminating objects within five arcseconds.

\begin{figure}
\begin{center}
\includegraphics[width=8.2cm,angle=0]{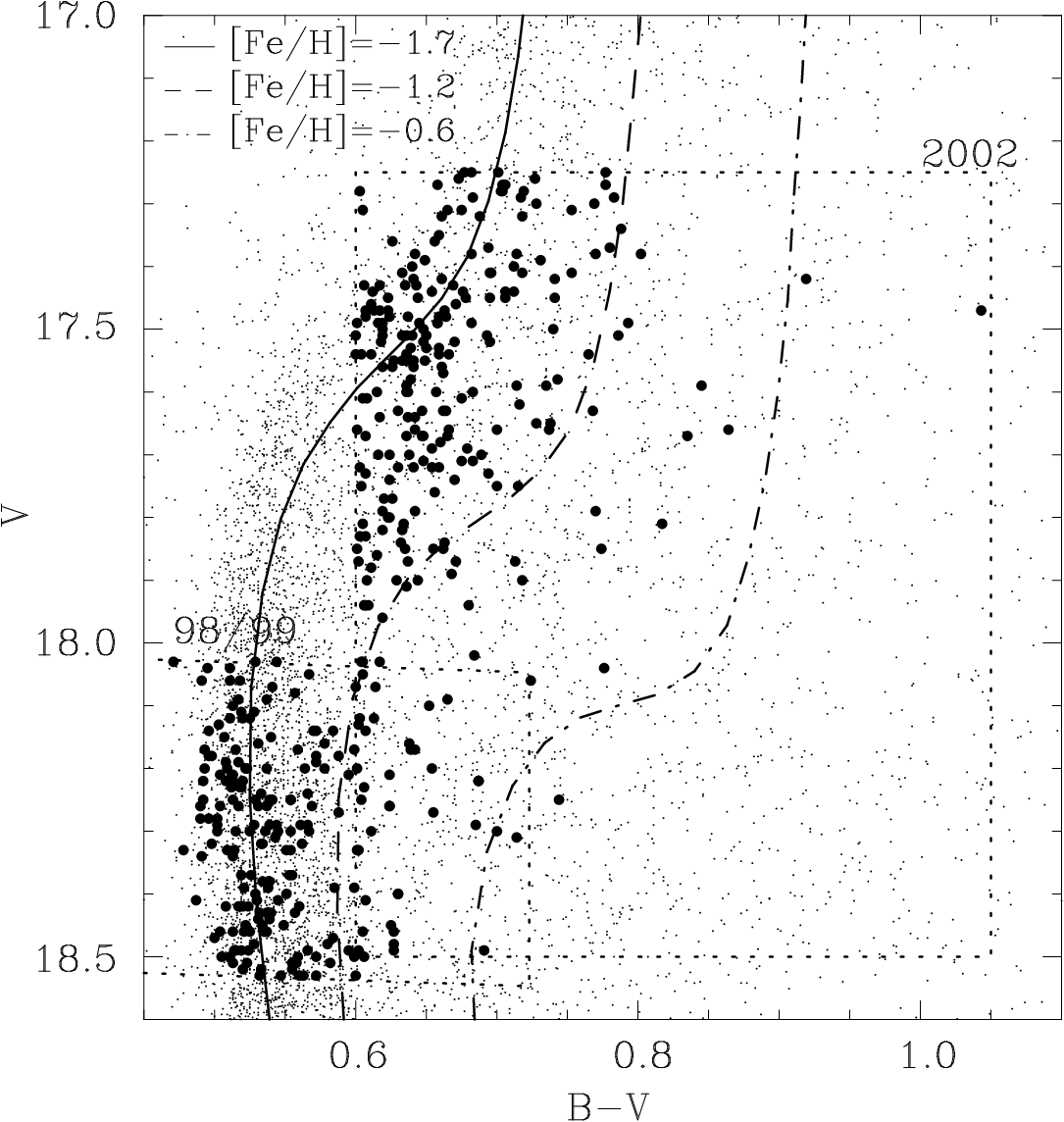}
\figcaption[f3.ps]{Color magnitude diagram for the members
found in the two observed samples, represented by large dots.  The
small dots are plotted to show where the bulk of the stars from the
photometry lie.  These have no membership information. The isochrones
are for [Fe/H]=--1.7,~--1.2,~--0.6 and each has an age of 13.5~Gyrs.
\label{cmd2} }
\end{center}
\end{figure}

\begin{figure}
\begin{center}
\includegraphics[width=8.2cm,angle=0]{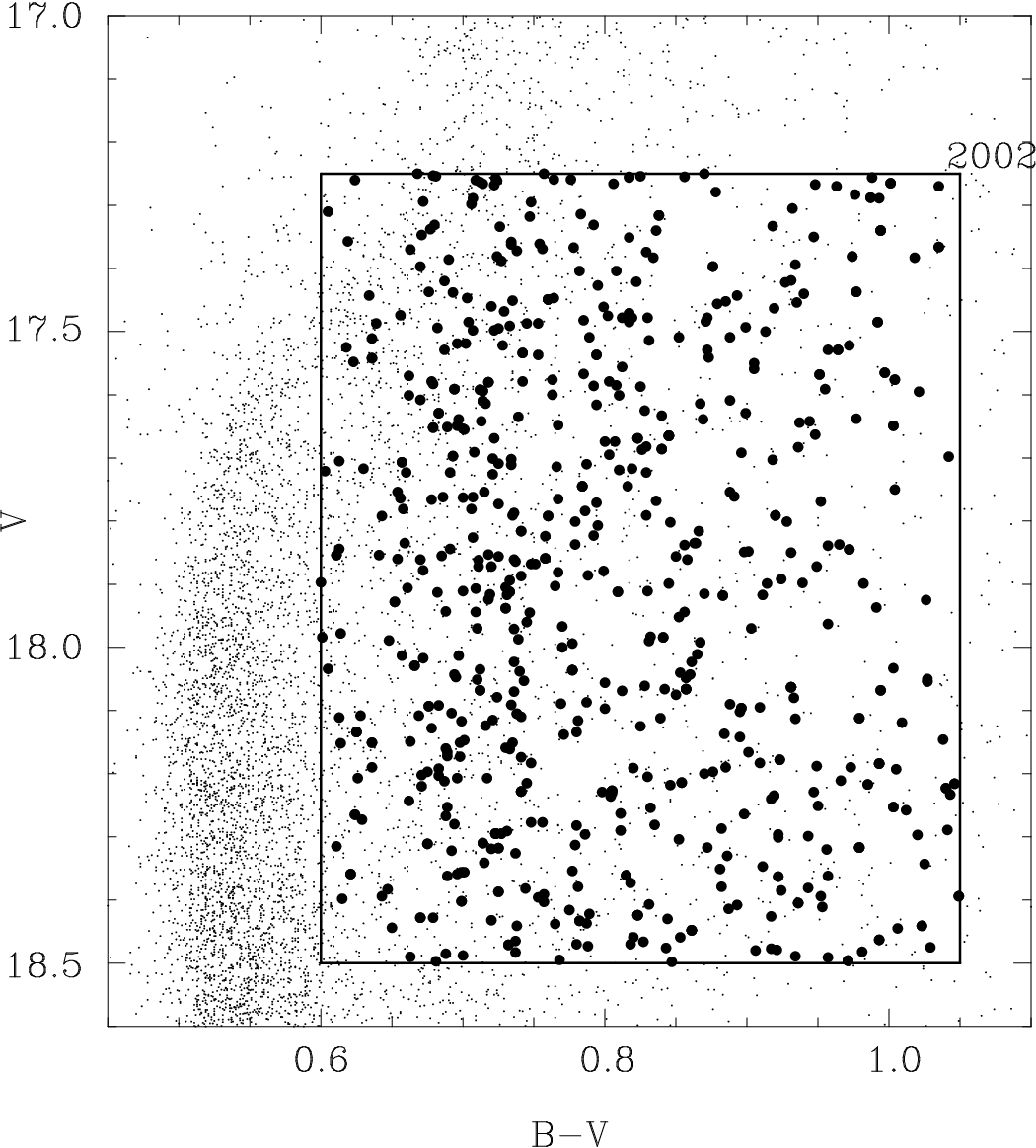}
\figcaption[f4.ps]{Color magnitude diagram for the
non-members in the 2002 observing run.
\label{nonmem} }
\end{center}
\end{figure}

These processes yielded a final sample of 442 members from $\sim$850
observed candidates near the turnoff in the CMD of $\omega$ Cen. A CMD
showing the members is shown in Figure~\ref{cmd2} (large dots). The
small dots represent the photometry as in Figure~\ref{cmd1}.  The
isochrones are the same as those in Figure~\ref{cmd1}.  This figure
shows that there are a number of very red, presumably metal-rich,
stars in {\wcen} at the turnoff region. Objects that were classified
as radial velocity non-members of the cluster in the 2002 sample are
shown in Figure~\ref{nonmem}.  This diagram shows that the candidates
were positioned fairly uniformly over the 2002 region on the CMD.  It
also makes clear that while we found no members in the lower right
corner of the box in Figure~\ref{cmd2}, candidates were observed
there.  Tables~\ref{tbl-2} and \ref{tbl-3} briefly lists the members
and non-members, respectively.  Both Tables are shown in their
entirety in the electronic version of The Astrophysical Journal.
Table~\ref{tbl-2} details the identification~(ID), right
ascension~(RA) and declination~(DEC) of each cluster member, along
with the V magnitude and color ($B-V$).  Column 6 lists the
heliocentric velocity.  Columns 7 and 8 list the determined
metallicity~([Fe/H]) and the error associated with
it~($\sigma_{[Fe/H]}$).  Columns 9 and 10 give the age assigned to
each star and its error.  Finally, column 11 informs the reader on
which run the star was observed, where 1 is for the 98/99 sample and 2
for 2002.  Table~\ref{tbl-3} gives the identification~(ID), right
ascension~(RA) and declination~(DEC) of the non-members, and
photometry information.  Column 6 lists the heliocentric velocity, and
the final column states on which observing run the star was observed.

\begin{figure}
\begin{center}
\includegraphics[width=7.2cm,angle=270]{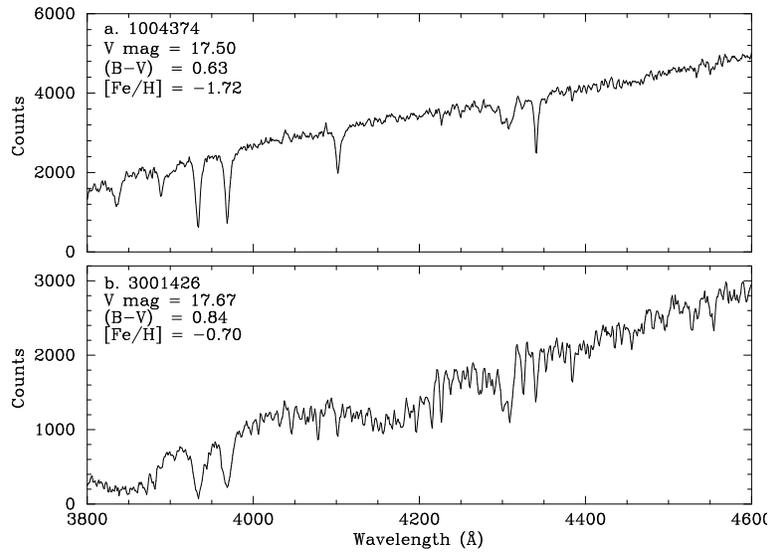}
\figcaption[f5.ps]{Spectra of two stars in our sample.  The
first, in the top panel, is a subgiant from the metal-poor population.
The second is of a subgiant from one of the more metal-rich
populations.  These spectra have a resolution of $\sim$2.4\AA.
Noticeable differences are the increased G band at $\sim$4300\AA, CN
at 3883\AA and 4215\AA\ and stronger metal lines in the more
metal-rich star.
\label{spectra} }
\end{center}
\end{figure}

Example spectra of a metal-poor subgiant, whose metallicity is
representative of the majority of the cluster population, and one of
the more unusual metal-rich member are shown in Figure~\ref{spectra}.
Noticeable differences between the two spectra are the increased G
band between 4250--4310\AA, CN at 3883\AA\ and 4215\AA\ and numerous
metal lines in the more metal-rich star.  A paper on the analysis of
the abundances of carbon, nitrogen, strontium and barium for all
stars, including the more peculiar objects, is in preparation.

Spectra with 2dF were also obtained for main sequences stars in four
other globular clusters, NGC~6397, M55, NGC~6752 and 47~Tuc.  These
clusters were observed in a similar manner to {\wcen}.  The data
obtained for the clusters were used here to test the reliability of
the metallicity calibration.  Observing dates and parameters used are
given in Table~\ref{tbl-4}.  Abundance standards were also observed
for calibrations during each 2dF cluster run.

\section{Metallicities} \label{met_sect}

\subsection{Abundance Calibration}

Metallicities were calculated using a combination of two methods
following \citet{bee99}. The first uses the Ca{\sc~ii}~K~line strength
(in the form of a pseudo-equivalent width) and $(B-V)_{0}$.  In order
to take into account the large range in K~line strength, three
different on-feature band widths were used.  These were 6, 12 and
18\AA\ and formed indices designated K6, K12 and K18 respectively.
The line bands and sidebands for these indices are defined as in
\citet{bee99}. The final index, K$'$, is given by:

\begin{equation}
\mbox{K}' = \left\{ \begin{array}{ll}
\mbox{K6}  & $K6 $\leq$ 2.0 \AA$          \\
\mbox{K12} & $K6 $>$ 2.0 \AA, K12 $\leq$ 5.0 \AA$   \\
\mbox{K18} & $K12 $>$ 5.0 \AA  $
\end{array} \right. 
\end{equation}

This metallicity determination becomes more uncertain at higher
metallicities due to the saturation of the
Ca{\sc~ii}~K~line. Another uncertainty arises from $(B-V)_{0}$, as
this color index may be affected by anomalous CN absorption not found
in the calibrating objects.  At fixed K$'$ and for a $(B-V)_{0}$
typical of the more metal-rich stars that show strong CN absorption, a
change of $(B-V)_{0}$ of $\pm$0.02 mag results in an abundance change
of $\pm0.08$~dex.

The second method used a technique known as the Auto-Correlation
Function (ACF) method which utilizes the strength of the metal lines
in the spectrum.  The wavelength range used in this technique is
$\lambda\lambda$4000--4285\AA\ excising the CN band and hydrogen lines
in the process (from 4166--4216\AA\ and 20\AA\ centered on
H$_\delta$). The main drawback of this calibration is that the spectra
need to have a sufficient signal-to-noise ratio (at least S/N~
$\approx$20) so that the method is not seriously affected by noise.

To quantify the usefulness of the two indices, tests were performed to
determine cutoff limits in S/N and in $(B-V)_{0}$ for which only K$'$,
only ACF, or the combination of both would be used.  The cutoff limits
were determined from spectra of 15 high proper motion stars chosen
from the lists of Giclas (see e.g.\ \citealt{clla96}) to cover the
range in metallicities found in {\wcen}.  Spectra of these objects,
which are all dwarfs, were obtained with ANU's 2.3m telescope at
Siding Spring Observatory and have a resolution of 1.2{\AA} {\sc fwhm}
and S/N$\sim$100.  The spectra were convolved to have the same
resolution as our 2dF sample and five different levels of noise were
introduced to the spectra to cover the S/N range in our sample.  The
convolved and noise-added spectra were then analyzed with our
metallicity determination technique, and the resulting abundances
compared, as a function of color and S/N, with the abundances that
result from applying our technique to the original higher resolution,
high S/N spectra.  As expected, objects with higher metallicities
showed larger errors when using the K$'$ index, while the metal-poor
stars had larger scatter when ACF was employed.

This procedure also found a systematic offset in the ACF abundance in
the sense that the abundances derived from the convolved noise-added
spectra were higher than those for the original spectra.  The offsets
were a function of metallicity, color and S/N, with bluer objects
having larger offsets than redder ones for a given S/N. Given the
similarity between the convolved noise-added spectra of the Giclas
dwarfs and those of our {\wcen} stars, we applied these corrections,
again as a function of metallicity, color and S/N, to the initial
{\wcen} star ACF metallicity determinations to account for this
systematic effect.  The corrections ranged in size from
--0.1~to~--0.3~dex.

We also analyzed 2dF observations of a number of field dwarfs on known
metallicities chosen from the lists of \citet{clla96} and
\citet{bee99}.  Approximately half of these spectra were obtained during
our cluster observing runs while the remainder were obtained from
other runs using the same instrumental setup.  Our K$'$ and ACF
abundances derived from these spectra generally agreed well with the
literature values: the mean differences were $\sim$0.1~dex for
[Fe/H]$<$--1.0, but somewhat higher ($\sim$0.2--0.3~dex) for the more
metal-rich objects.

\begin{figure}
\begin{center}
\includegraphics[width=6.2cm,angle=0]{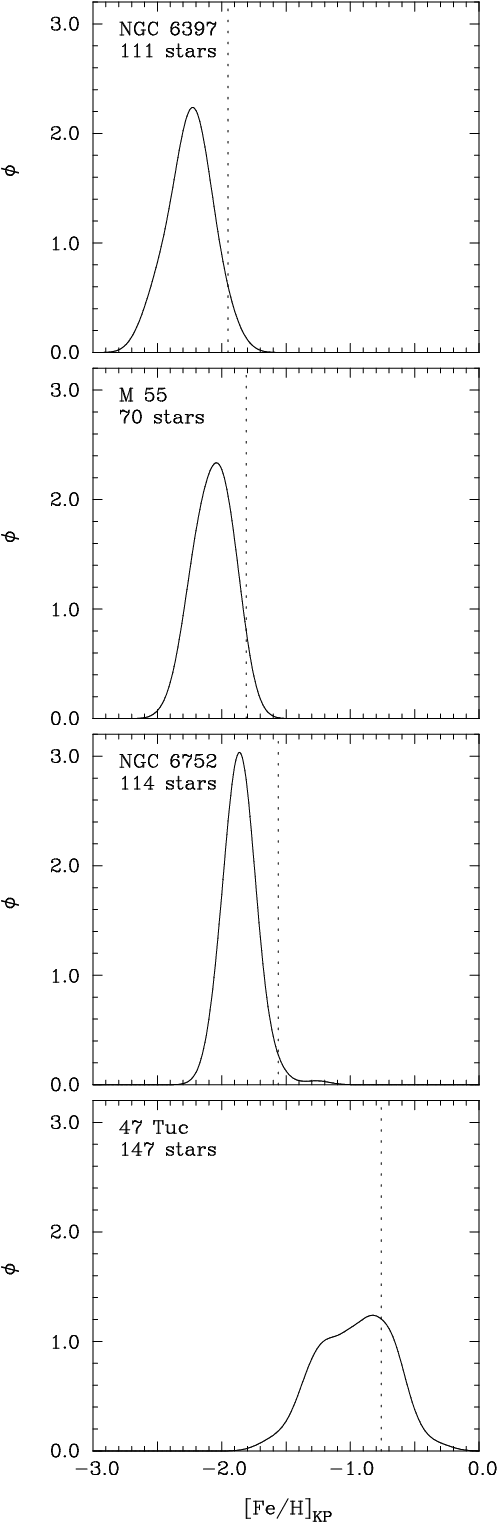}
\figcaption[f6.ps]{K$'$ metallicity generalized histograms
for the members of the calibrating clusters NGC~6397, M55, NGC~6752
and 47~ Tuc.  The dotted line in each panel indicates the accepted
[Fe/H] value \citep{har96}, which is consistently of higher
metallicity than the mean [Fe/H] determined here.
\label{ogckp}}
\end{center}
\end{figure}

\begin{figure}
\begin{center}
\includegraphics[width=6.2cm,angle=0]{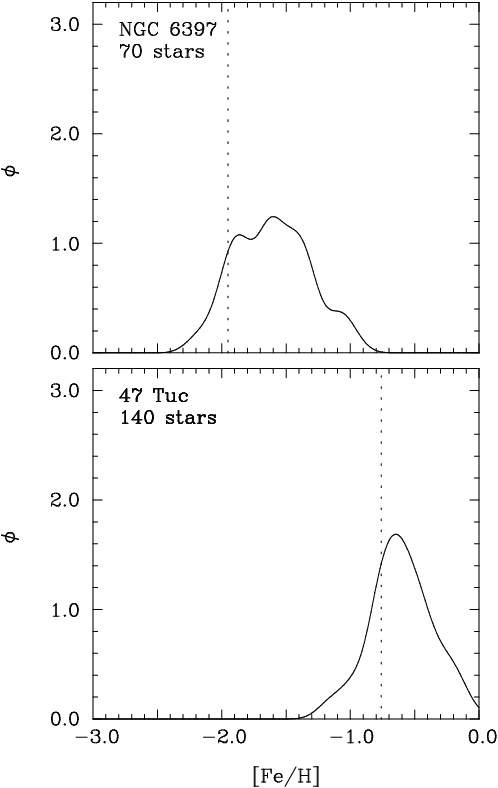}
\figcaption[f7.ps]{ACF metallicity histograms for NGC~6397
and 47~Tuc.  Again, the dotted line shows the accepted [Fe/H] for the
cluster. Note that this line is more metal-poor than the mean of
each distribution.
\label{ogcacf}}
\end{center}
\end{figure}

The existence of 2dF spectra for substantial samples of main sequence
stars in four globular clusters (NGC~6397, M55, NGC~6752 and 47~Tuc)
allows a further investigation of our abundance determinations.  These
spectra have similar S/N to our {\wcen} sample. We applied our
technique to the cluster main sequence star spectra and the resulting
abundance histograms for [Fe/H]$_{K'}$ and [Fe/H]$_{A}$ are shown in
Figures~\ref{ogckp} and \ref{ogcacf} respectively.  In the case of
[Fe/H]$_{K'}$, we find that the mean abundances for the clusters from
our technique are systematically low by 0.2 to 0.3 dex compared to the
accepted metallicities \citep{har96}.  We note also that the
distribution for 47~Tuc is considerably broader than for the other
three clusters.  This results from the saturation of the
Ca{\sc~ii}~K~line at higher in metallicities, which, in turn,
causes a larger uncertainty in the final [Fe/H]$_{K'}$.  Further, we
verified that the possible bimodality of the 47~Tuc abundance
distribution in Figure \ref{ogckp} is not the results of the known
bimodality in CN strengths on the cluster main sequence.

For the ACF metallicity, despite use of the offsets defined from the
analysis of the Giclas star spectra, the values for NGC~6397 and
47~Tuc are higher than the accepted values by $\sim$0.2~dex (no
[Fe/H]$_{A}$ values were derived for the M55 and NGC~6752 stars as the
spectra generally do not possess sufficient counts to apply the
technique). We note also that the width of the NGC~6397 [Fe/H]$_A$
distribution is quite broad. This is a consequence of the relatively
low sensitivity of the ACF method at low abundance.

It is not clear why the clusters give systematic offsets in the
[Fe/H]$_{K'}$ and [Fe/H]$_{A}$ abundances while the standard stars do
not, although it may be that we simply do not have enough standard
objects to thoroughly test for any offsets.  Our preference is to use
the cluster data rather than that of the standard stars given the
much larger sample sizes in each cluster at a given [Fe/H] (hundreds
versus a few). It was also considered better to use objects that were
similar in magnitude and color and were observed and reduced in the
same manner as the {\wcen} stars.

\begin{figure}
\begin{center}
\includegraphics[width=6.2cm,angle=0]{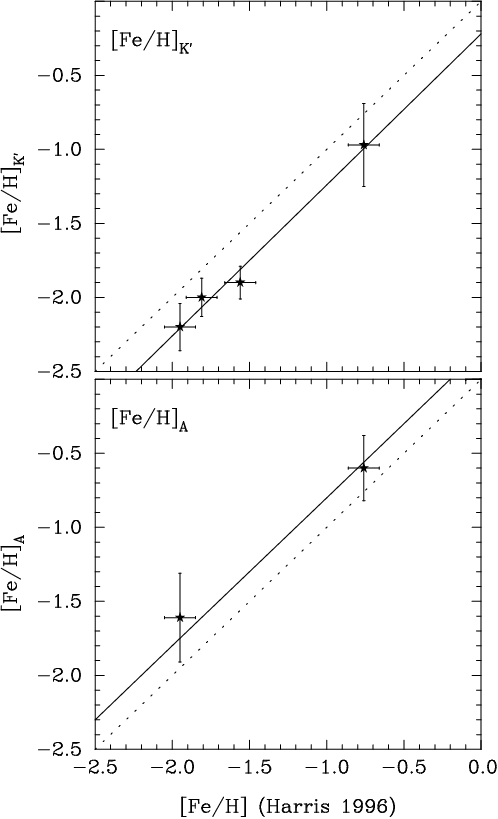}
\figcaption[f8.ps]{Comparison of the mean  [Fe/H]
determined from the cluster metallicity distributions with the accepted [Fe/H]
\citep{har96}.  The upper panel is for the K$'$ calibration, while the
bottom shows the results from the ACF calibration.  The dotted line in
each panel indicates the 1:1~line.  The solid line in the upper panel
is the least-squares fit to the data, and in the lower panel
represents the 0.2~dex offset to the 1:1~line.
\label{ogcfc}}
\end{center}
\end{figure}

Comparisons between the mean K$'$ metallicities for each cluster and
the accepted [Fe/H] values are shown in Figure~\ref{ogcfc}, along with
a similar comparison for the ACF metallicities.  The median of the
distributions for 47~Tuc and for NGC~6397 ([Fe/H]$_A$) were used here.
In these plots the dotted lines are 1:1 relations.  In the upper
panel, the solid line is the least-squares fit to the data, while the
lower panel's solid line is an offset of 0.2~dex to the 1:1 line.
Corrections to the K$'$ metallicities were constructed based on a
linear fit from the calibrating clusters and were applied to the
{\wcen} data.  The ACF correction used was an 0.2~dex offset from the
1:1 line.  Fitting a straight line to the ACF data may not give an
accurate correction as there are only two data points and to err on
the side of caution, we instead used an offset.

\begin{equation} 
\mbox{[Fe/H]}_{\mbox{K}'\mbox{c}} =  \frac{\mbox{[Fe/H]}_{\mbox{K}'}+0.17}{1.06}
\end{equation}

\begin{equation} 
\mbox{[Fe/H]}_{\mbox{Ac}} =  {\mbox{[Fe/H]}_{\mbox{A}}-0.2}
\end{equation}

The K$'$ calibration is more reliable for metal-poor stars due to the
saturation of the Ca{\sc~ii}~K~line at higher metallicities. The
ACF method, on the other hand, is more reliable for the metal-rich
objects due to the loss of sensitivity at lower abundances.  Therefore
limits were put in place at metal-poor and metal-rich ends of our
metallicity range to use only the method that suited best. The final
metallicity was given by:

\begin{equation} 
\mbox{[Fe/H]} = \left\{ \begin{array}{ll}
\mbox{[Fe/H]}_{\mbox{K}'\mbox{c}}  & \mbox{[Fe/H]}_{\mbox{K}'\mbox{c}}  \mbox{ or [Fe/H]} _{\mbox{Ac}} \leq -2.0\\
\mbox{[Fe/H]}_{\mbox{Ac}} & \mbox{[Fe/H]}_{\mbox{K}'\mbox{c}} \mbox{ or [Fe/H]} _{\mbox{Ac}} \geq -0.8\\
\langle\mbox{[Fe/H]}\rangle & \mbox{otherwise} \\
\end{array} \right. 
\end{equation}

The weighted mean of the two metallicities was calculated using:

\begin{equation} 
\langle\mbox{[Fe/H]}\rangle = \frac{\frac{\mbox{[Fe/H]}_{\mbox{K}'\mbox{c}}}{\sigma^{2}_{\mbox{K}'\mbox{c}}} + \frac{\mbox{[Fe/H]}_{\mbox{Ac}}}{\sigma^{2}_{\mbox{Ac}}}}
{\frac{1}{\sigma^{2}_{\mbox{K}'\mbox{c}}} + \frac{1}{\sigma^{2}_{\mbox{Ac}}}}
\end{equation}

The error estimates associated with the initial K$'$ and ACF
determinations were derived from two sources. The first is from the
Beers formulation itself, where an error estimate is assigned for the
K$'$ and ACF metallicities individually as described in \citet{bee99}.
The second source comes from the fact that the S/N of our data is
lower than the average S/N of the spectra used in the Beers
calibration. This takes account of the fact that the ACF metallicities
are impacted more by the noise level in the spectra, which is
particularly the case for the metal-poor stars.  Using the procedure
to correct the metallicities described earlier, errors were assigned
to the metallicities based on the stars color, metallicity and S/N
($\sigma\leq$0.2~dex).  These two sources were added quadratically to
give the error estimate.

An independent estimate of the errors used the data from the
calibrating clusters.  Two of the calibrating clusters have
metallicities which are approximately equal to our outer boundaries at
[Fe/H]=--2.0 and~--0.8.  The standard deviation was calculated for the
metallicity distribution for these clusters and adopted as the error
($\sigma$) for the {\wcen} stars at the metallicity of the cluster.
For NGC~6397 and 47~Tuc, $\sigma_{K'}$=0.14 and 0.29, and
$\sigma_{ACF}$=0.30 and 0.27, respectively.  Linear interpolation was
used between the two metallicities in order to assign errors at all
metallicities between our boundary limits.  This technique was
performed for both ACF and K$'$ abundances.  At lower metallicities
the standard deviation for K$'$ was lower than that for ACF.
Conversely the ACF standard deviation was lower at higher
metallicities than K$'$.

For the K$'$ abundance error, the first estimate was higher than the
second by $\leq$0.1 dex. The opposite was true, however, for the ACF
metallicity error by the same amount.  For both K$'_{c}$ and ACF$_c$
the final adopted error was taken as the average of the two separate
estimates, while the overall error associated with a given metallicity
was taken as the quadratic sum:

\begin{equation} 
\frac{1}{\sigma_{\mbox{F}}^2} = \left( \frac{1}{\sigma^{2}_{\mbox{K}'\mbox{c}}} + \frac{1}{\sigma^{2}_{\mbox{Ac}}} \right)
\end{equation}

\subsection{Metallicity Distribution}

\begin{figure}
\begin{center}
\includegraphics[width=6.2cm,angle=0]{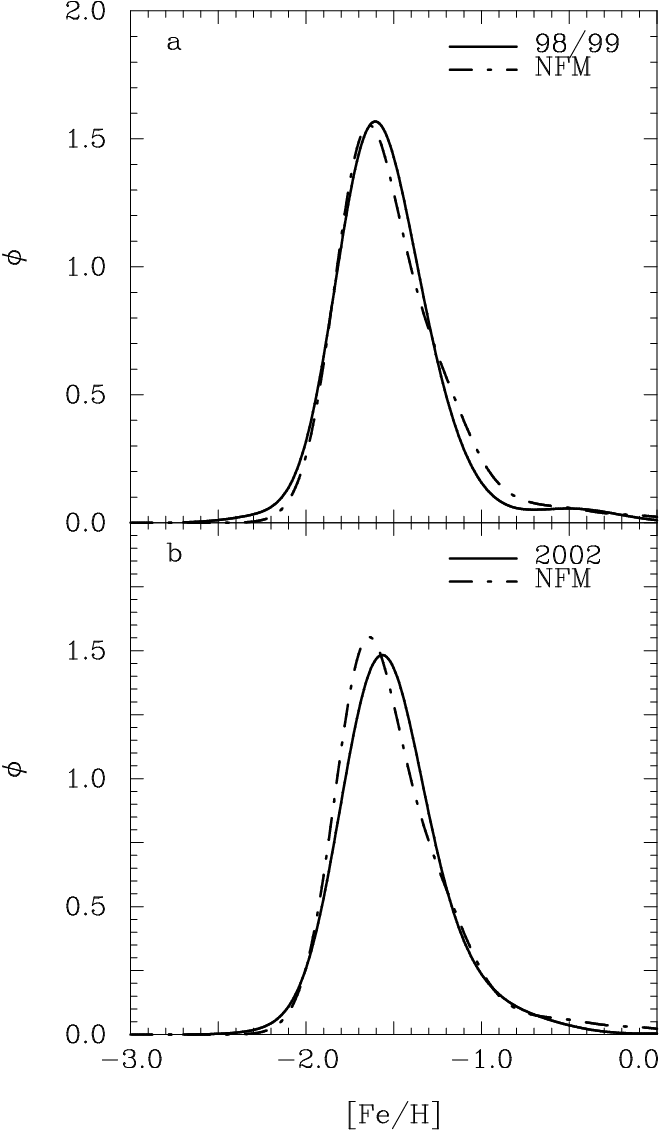}
\figcaption[f9.ps]{Generalized histograms of the
metallicity distribution for the 98/99 (panel~a) and 2002 (panel~b)
samples (gaussian kernal of $\sigma$=0.15--0.20, based on the
individual errors for each metallicity).  For comparison, the
\citet{nfm96} red giant branch distribution is also plotted
($\sigma$=0.14).
\label{hist}}
\end{center}
\end{figure}

Figure~\ref{hist} shows the resulting metallicity distributions
obtained for the 98/99 (panel~a) and 2002 samples (panel~b).  The
distributions are compared with that found for the RGB of the cluster,
taken from \citetalias{nfm96}. To convert their [Ca/H] distribution to
[Fe/H], [Ca/Fe] was assumed to be 0.3 for [Fe/H]$\leq$--1.0
\citep{scl95, pan00}, declining linearly from [Ca/Fe]=0.3 at
[Fe/H]=--1.0 to 0.0 for [Fe/H]=0.  The distributions have been
normalized by area and the RGB distribution has been convolved with a
wider gaussian kernal ($\sigma$=0.14) than in \citetalias{nfm96} due
to the larger errors associated with our metallicities.  The
metallicity errors for our sample are $\sim$0.15--0.2~dex, compared
with 0.05~dex for that of the \citetalias{nfm96}.  The generalized
histograms for the 98/99 and 2002 samples utilize the individual
$\sigma$ associated with each metallicity.

The 98/99 spectra were of hotter and fainter objects. The range of S/N
for the 2002 sample was $\sim$30--70, while for the 98/99 spectra it
was $\sim$20--40.  The lower S/N for the 98/99 sample and the fact
that the majority had metallicities calculated using the K$'$
calibration instead of the combination of the ACF and K$'$
calibrations made their metallicity determinations slightly more
uncertain. Since it is the 2002 sample, in particular the brighter
stars at the turnoff, that gives us the most information about the age
range in the cluster the larger errors in abundance on the
main-sequence stars are not a great concern for the age
determinations.

The data from our two sub-samples were not combined due to the 2002
set being incomplete at the metal-poor end and biased towards the
metal-rich populations.  The 98/99 sample is unbiased with respect to
the distribution of members in the CMD (except for a small number of
stars with $B-V>0.72$), while the 2002 one is biased against the
metal-poor sample.  This can be seen in the offset between the 2002
sample and the \citetalias{nfm96} data, as no corrections have been
made for selection effects.  There have also been no evolutionary
corrections made to the distributions, though these are likely to be
minor.

The distributions have a steep rise at [Fe/H]=--1.7, with tails to
higher metallicities.  We find 25\% (44/174) and 15\% (39/254) of
stars with metallicities [Fe/H]$<$--1.7 for the 98/99 and 2002
samples, respectively.  Stars with [Fe/H]$>$--1.0 account for 4\%
(11/254) for the 2002 sample and 5\% (8/174) for the 98/99 sample.
In the NFM data there are 25\% of stars with [Fe/H]$<$--1.7 and 6\%
of stars with [Fe/H]$>$--1.0.  The reader should note that the NFM
sample covers the whole region of the cluster, while our sample covers
the outer region between 15 and 25~arcminutes.  We conclude that the
metallicity range found for the turnoff region is qualitatively
similar to that found for the giant branch.

A Kolmogorov-Smirnov two-sample test was performed on the 98/99 and
\citetalias{nfm96} distributions.  The null hypothesis was that the
two samples came from the same distribution.  This test was also
repeated using the data sets from \citet{sk96} in place of that of
\citetalias{nfm96}.  Their data for {\wcen} comprise two groups, one
of subgiant branch objects and the other of red giant branch
stars. These three sets of data were tested against our 98/99 data
separately.  For the 98/99 data, we found that the null hypothesis
could not be rejected.  Not surprisingly given the biased selection,
the 2002 data set showed a different result, and the null hypothesis
was rejected for each of the three tests. 

\begin{figure}
\begin{center}
\includegraphics[width=12.7cm,angle=0]{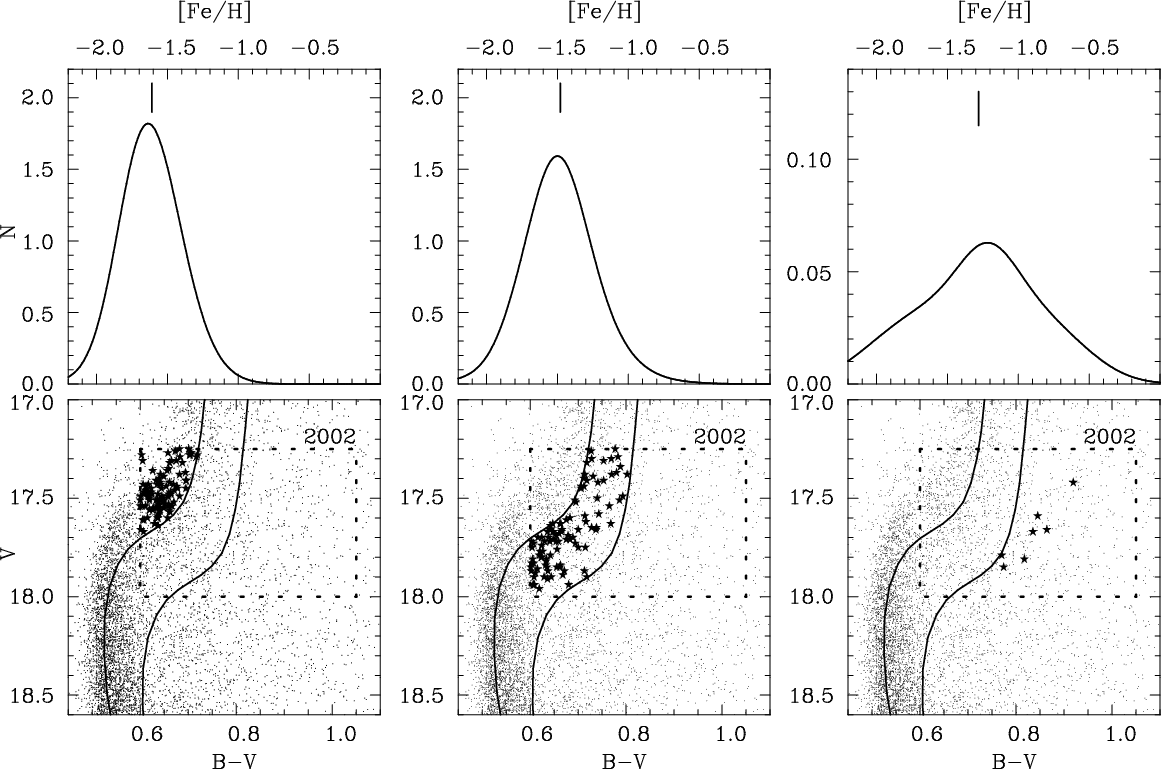}
\figcaption[f10.ps]{CMD split into three
different areas, designated by the solid lines which were isochrones
of the two metal-poor populations but shifted in color and luminosity.
Generalized histograms were constructed using the data for the corresponding
groups.  The histograms show the metallicity distributions for those
areas on the CMDs.
\label{cmdhist}}
\end{center}
\end{figure}

To check the accuracy of the metallicities the members falling in the
2002 turnoff box were separated into three groups based on their
photometry shown in the lower panels of Figure \ref{cmdhist}.This
figure illustrates the differences in metallicity as a function of
position on the CMD.  The solid lines indicate where the regions of
interest lie.  These lines are based on an isochrone where the first
(from the left) has parameters of [Fe/H]=--1.2, age=13.5~Gyrs, and
offset in $V$ by --0.14~mag, and in $(B-V)$ by -0.062~mag. The second
solid line is the same isochrone but offset by $V$=0.12~mag and
$(B-V)$=0.018~mag.

Corresponding metallicity histograms for each group were plotted shown
in the upper panels.  The first group has a mean
[Fe/H]=--1.61$\pm$0.13, the second [Fe/H]=--1.48$\pm$0.17.  The third
group has a mean [Fe/H]=--1.28, but note the small number of objects
in this group. The errors in abundance for the third group are large
(0.3~dex), evident by the large width of the histogram.

\section{Ages} \label{ages_sect}

When determining the ages using theoretical isochrones, it is best to
use the turnoff region since this is where the isochrones are more
sensitive to age variations. For the present investigation only the
members at the turnoff with $V\leq$18 were used, and these stars
came from the 2002 sample.

Two methods were used to calculate the age range of the cluster. The
first involved assigning individual ages to each star based on its
position on the CMD and metallicity using theoretical isochrones. The
second method involved the construction of synthetic CMDs from a
specified metallicity distribution, age range and theoretical
isochrones, and comparison between synthetic and observed CMDs.

The isochrones used were the Yonsei-Yale (Y$^2$) isochrones
\citep{yi01, kim02}. These isochrones permit interpolation between
age, metallicity and alpha elemental abundance to generate the
required isochrone. A grid of isochrones was used which span the
metallicity range --2.6$<$[Fe/H]$<$0.3 in 0.05 dex increments. For
each metallicity there were 34 isochrones for ages 2--19~Gyrs in
0.5~Gyr steps. Alpha enhancement was taken to be constant
([$\alpha$/Fe]=0.3) for [Fe/H]$\leq$--1.0, and declining linearly
for higher [Fe/H] until it reached the solar value at [Fe/H]=0.

The RGB metallicity distribution from \citetalias{nfm96} was used as
the input into all simulations when requiring synthetic CMDs in the
following sections.  This distribution was shown in the previous
section to be similar to the one found on the main sequence.  As the
RGB distribution is for [Ca/H] rather than [Fe/H], it was scaled using
constant [Ca/Fe]=0.3 for [Fe/H]$<$--1.0 and linearly decreasing
[Ca/Fe]~to~0.0 at [Fe/H]=0.0 for [Fe/H] greater than --1.0.

\subsection{Method 1: Assigning Individual Ages to Stars} \label{meth1_sect}

To assign an age to each star, its metallicity was used to select the
nearest isochrone in our grid. The isochrones with this metallicity
but with differing ages were then compared to the star's $(B-V)_0$ and
M$_{V}$ on the CMD to find the one closest. Usually a star's position
did not fall directly on one isochrone and linear interpolation in
color was performed between the two closest ones to determine its age.

An error associated with the age was obtained using the errors in $B-V$
and metallicity. The errors associated with the individual V magnitude
contributes a very small amount to the final error and was therefore
ignored here. The age calculation was repeated for a positive and
negative change in color using the values given in Table~\ref{tbl-1}.
Similarly the metallicity was modified using $\pm$1$\sigma$ errors to
obtain the corresponding error in age.  The range in age determined by
the metallicity errors and that determined by the photometry errors
were quadratically summed and halved to give the final estimate of the
age error.

 Total errors in the age calculation were up to $\pm$4~Gyrs for stars
below the turnoff where the isochrones are close together, and up to
$\pm$2~Gyrs for objects above the turnoff.  There were some stars that
did not fit any of the isochrones in the grid for their
metallicity. These were given the maximum (or minimum) age in the
range i.e. 19~(or 2)~Gyrs and represented~$\sim$7\% of the sample. An
age of 19~Gyrs for a object in a globular cluster is not believable,
nor is one of 2~Gyrs. This discrepancy is probably due to errors in
the photometry or calculated metallicity.

\begin{figure}
\begin{center}
\includegraphics[width=8.2cm,angle=0]{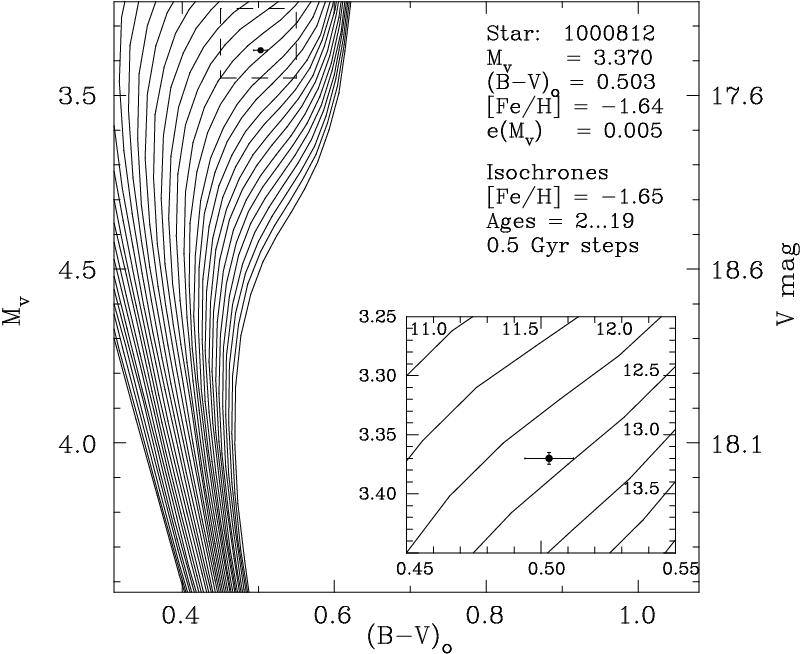}
\figcaption[f11.ps]{Age-metallicity diagram for the
$\omega$ Cen sample, showing only the turnoff stars, as this is the
area of the CMD that is most sensitive to age.
\label{am1} }
\end{center}
\end{figure}

Figure~\ref{am1} shows the age-metallicity diagram (AMD) resulting
from this method. This plot shows only those stars that lie at the
turnoff region of the CMD with $V<$18.0, where the metallicity-age
degeneracy is best broken.  It shows that there is an age-metallicity
relation in the cluster with the more metal-rich stars being
younger. A line of best fit to the data was drawn, by eye, excluding
stars at the upper age limit object (at 19~Gyrs) and is shown as the
solid line. This fit gives an age range of $\sim$5~Gyrs between
[Fe/H]=--1.7 and~--0.6.  As discussed below, however, this relation is
influenced by correlated errors.

To test the age range found above we performed Monte-Carlo simulations
of a population which had the metallicity distribution taken from
\citetalias{nfm96}.  The synthetic population of stars occupied the
same position on the CMD as the {\wcen} turnoff stars in our sample,
and had the same sample size.  Four different age ranges were
considered --- ~0,~2,~4 and~6~Gyrs between metallicities [Fe/H]=--1.7
and~--0.6, with a linear interpolation in age between these
metallicities.  The oldest population in each case was assigned an age
of 13.5~Gyrs and [Fe/H]=--1.7.  For example, the ages for the 2~Gyr
age range simulation was calculated as follows:

\begin{equation} 
\mbox{Age}_{\star} =  \left\{ \begin{array}{ll}
13.5  & \mbox{[Fe/H]}_{\star} \leq -1.7 \\
-1.82 \mbox{[Fe/H]}_{\star} + 10.41 & -1.7<\mbox{[Fe/H]}_{\star}<-0.6\\
11.5  & \mbox{[Fe/H]}_{\star} \geq -0.6 \\
\end{array} \right. 
\end{equation}

Photometric errors were included that were representative of the
observed sample (see Table~\ref{tbl-1}). We also included an error on
the abundance determination ($\sigma$=0.15 dex) when simulating the
populations. We then determined ages for all stars in each of the
populations in the same manner as was done for the observed data.  The
goal was to test how well we could recover the input parameters given
the errors on photometry and metallicity.

\begin{figure}
\begin{center}
\includegraphics[width=6cm,angle=0]{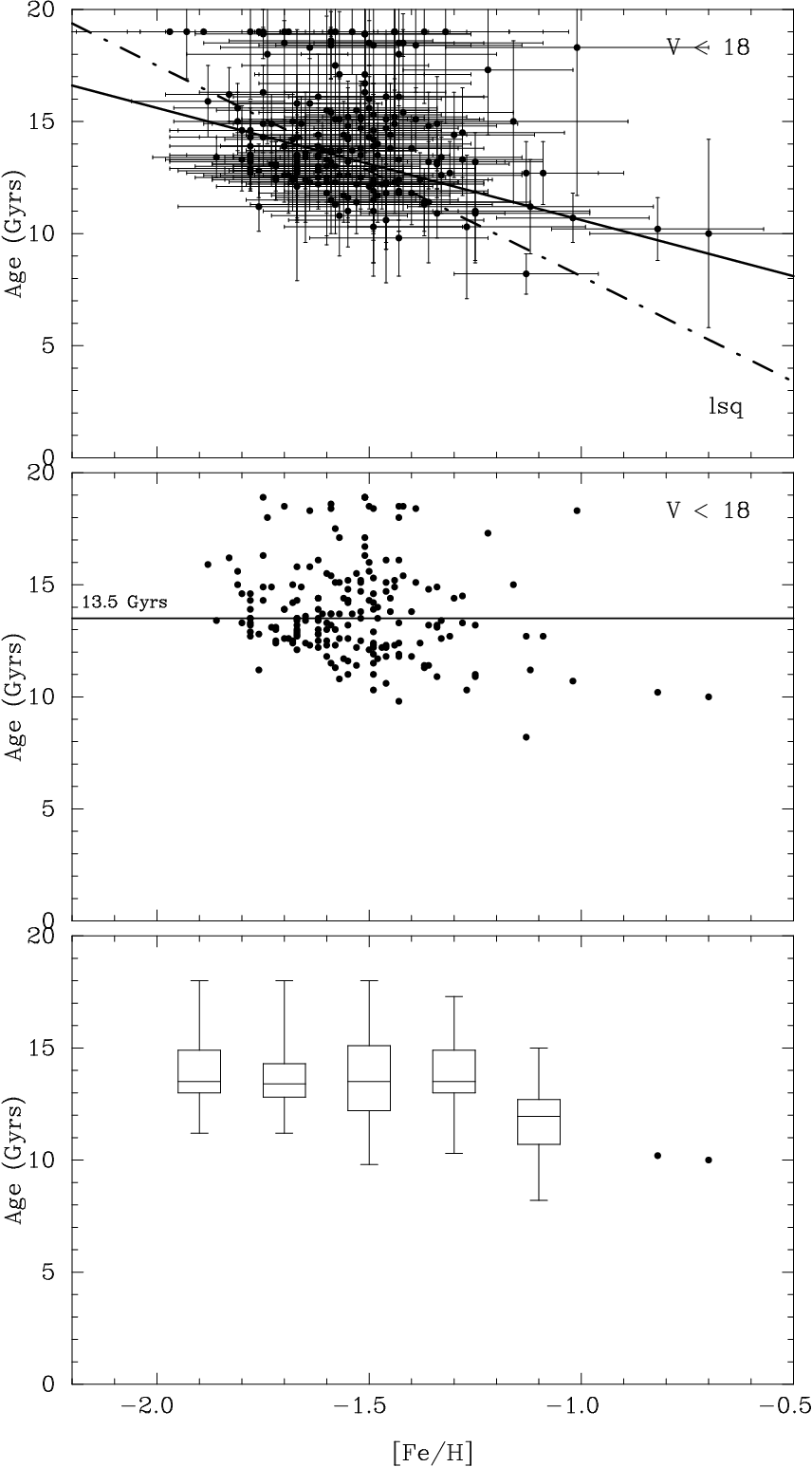}
\figcaption[f12.ps]{Age-metallicity diagrams for the
simulations. Panel~a shows the simulation with an input age of 0~Gyrs.
Panels~b,~c and~d show simulations with input ages of 2, 4 and 6~Gyrs,
respectively. The input metallicity range comes from \citet{nfm96}.
The dotted line in each panel represents the age-metallicity relation
(Sim.~AR) adopted for each simulation.  The solid line is the
least-squared fit to the data giving a calculated age range
(Calc.~AR).  See text (\S \ref{meth1_sect}) for more details.
\label{agemod} }
\end{center}
\end{figure}

The resulting AMDs are shown in Figure~\ref{agemod}.  In these plots,
each point represents a simulated star, and was assigned an age
depending on the input parameters.  Simulation 1 has no age range,
simulations 2, 3 and 4 have age ranges of 2, 4 and 6~Gyrs,
respectively, between metallicities [Fe/H]=--1.7 and~--0.6.  Any
stars that have abundances beyond those values have the maximum or
minimum ages assigned to them.  The dot-dash line in each plot
indicates the input age-metallicity relation before any photometric or
metallicity errors were included.  The solid line is the least
squares fit to the data, which takes errors in both coordinates into
consideration.

\begin{figure}
\begin{center}
\includegraphics[width=6.2cm,angle=0]{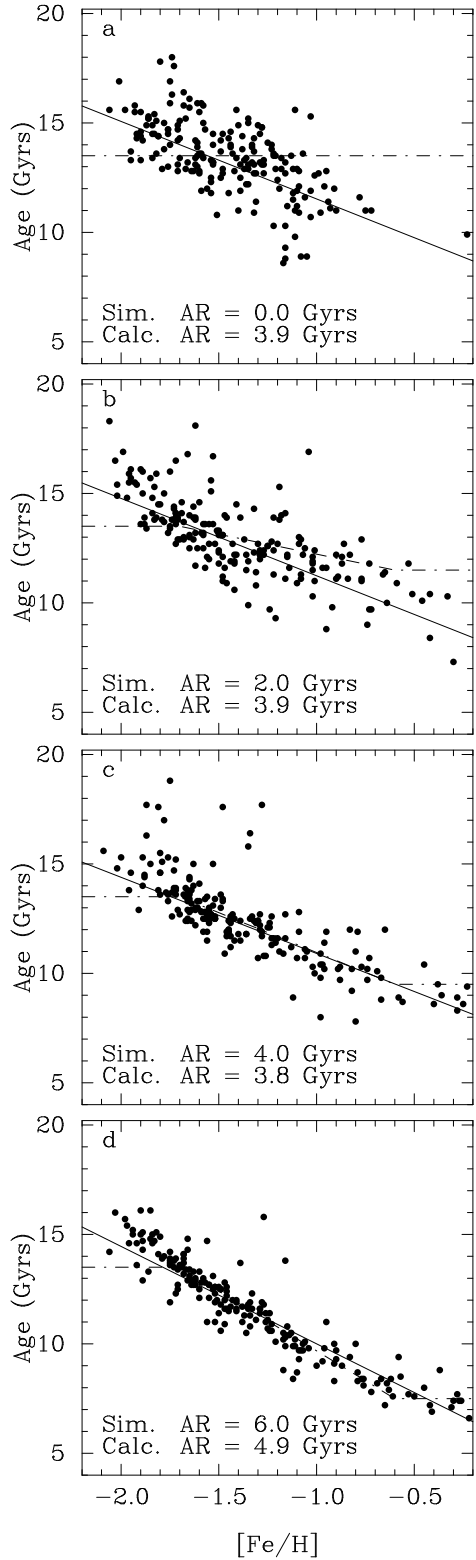}
\figcaption[f13.ps]{Observed age-metallicity diagram for
{\wcen} for comparison with the simulations in Figure
\ref{agemod}.  This is the same as Figure~\ref{am1} without the error
bars or 19~Gyr old stars.
\label{am2} }
\end{center}
\end{figure}

The first thing to note in Figure~\ref{agemod} is that errors induce
an age-metallicity relation.  Panel~a, which has an input age range of
0~Gyrs shows an apparent age range of 3.9~Gyrs.  As the input age
range becomes larger, the calculated age range comes more into line
with it.  The simulations with 4 and 6 Gyr age spreads do not
accurately reproduce the observed AMD for {\wcen} which can be seen by
comparing Figure~\ref{agemod} with the observations in
Figure~\ref{am2} (the latter is similar to Figure~\ref{am1} but
without the error bars).  For these simulations, the correlation
between age and metallicity is too tight when compared with that of
the {\wcen} plot.  The two simulations with age ranges of 0 and 2~Gyrs
show a scatter that is similar to that of {\wcen}, and have similar
slopes.  This shows that while the observational data indicate an age
range of $\sim$5~Gyrs, this figure drops considerably when the errors
in metallicity and photometry are taken into account.  From this we
conclude that the age range in {\wcen} lies between 0 and 2~Gyrs.  To
summarize: errors in metallicity and photometry have the potential to
induce an unreal age-metallicity relation, or to make a small one
appear larger.

These simulations also show the evolutionary effects for different
metallicities and ages.  Simulations with higher age ranges have more
metal-rich objects than those with no or small age ranges.  This is
due to our choice of turnoff box and to metal-rich young stars
spending a longer period within the box than the older metal-rich
objects.  As only a small number of metal-rich objects were found in
our observed sample, the above result strengthens the case for a low
age range within the cluster.

\begin{figure}
\begin{center}
\includegraphics[width=6.2cm,angle=0]{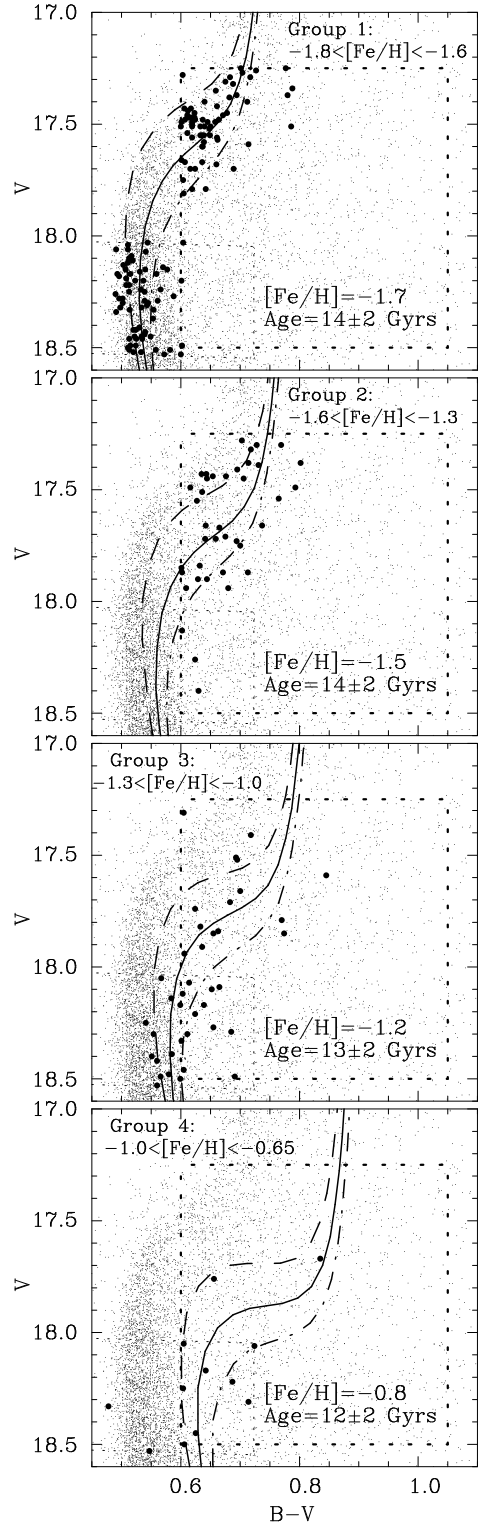}
\figcaption[f14.ps]{Metallicity cuts were made in the data,
shown at the top of each panel, and isochrones with metallicities
corresponding to the mean metallicity in each group were plotted with
ages that best fit the data.  This figure confirms the age range found
in \S\ref{meth1_sect}.
\label{cmdage}}
\end{center}
\end{figure}

A further check of the age range found is to plot the data within
metallicity ranges on the CMD and fit isochrones of appropriate
metallicities and ages (all with [$\alpha$/Fe]=0.3), shown in Figure
\ref{cmdage}.  Four metallicity bins were chosen.  The first had a
mean metallicity of [Fe/H]=--1.7, and an appropriate isochrone with
age 14~Gyrs was found to best fit the data.  Isochrones with age of 12
and 16~Gyrs were also plotted as dotted lines to show the the
approximate range in ages covered by the data.  Group 2 had a mean
metallicity of [Fe/H]=--1.5.  There is a broader spread in
metallicities in this group, accounting for the members that are not
encompassed by the isochrones plotted.  The third group had a mean
metallicity of [Fe/H]=--1.2.  Isochrones of this metallicity and
ages~=~13$\pm$2~Gyrs were found to fit the data, although there are a
few outliers in this group.  The mean [Fe/H]=--0.8 of the fourth group
was also used to plot isochrones with ages~=~12$\pm$2 Gyrs.  This
figure indicates that there is indeed a age range in the cluster.

\subsection{Helium abundance variations} \label{he_sect}

It has been suggested that large helium variations
($\Delta$Y$\sim$0.12) play a key role in understanding the bimodality
of the MS \citep{bed04, nor04, pio04}. It is therefore important to
see what effect helium variations have at the turnoff region of the
CMD and the ages calculated for our sample.  Unfortunately the Y$^{2}$
isochrones (as well as most other sets of published isochrones) do not
present results for sufficiently large ranges of helium for a given Z.
Ones that do cover the required range in helium and Z to date are the
Revised Yale Isochrones (RYI) \citep{gdk87}.  While these models do
not contain the most up-to-date physics, they are adequate to show the
relative effects of helium variations.

To test the effect of helium we assumed our sample had two populations
-- the first had [Fe/H]$\leq$--1.45 while the second had
[Fe/H]$>$--1.45.  The standard value of Y~(=0.23) was applied to the
first population, and helium for the second population would then be
0.35 (for $\Delta$Y=0.12). 

Although we wished to use Y=0.23 and Y=0.35 (for $\Delta$Y=0.12) for
the first and second groups, respectively, the RYI are incomplete
above Y=0.3 and ages are not able to be calculated at the turnoff.
Therefore values of Y=0.2 and~0.3 were used to show the relative
effect (using $\Delta$Y=0.10).  To test the effect of helium on the
ages, two sets of isochrones were used.  Both covered a metallicity
range of --2.6$<$[Fe/H]$<$--0.5 and an age range of 6--20~Gyrs.  All
the isochrones in the first set had Y=0.2.  For isochrones in the
second set that had [Fe/H]$\leq$--1.45, helium was taken to be~0.2,
and for those with [Fe/H]$>$--1.45, Y=0.3.  Ages were calculated using
these isochrones for each of our members in the cluster in the same
manner as described is \S\ref{meth1_sect}.

\begin{figure}
\begin{center}
\includegraphics[width=6.2cm,angle=0]{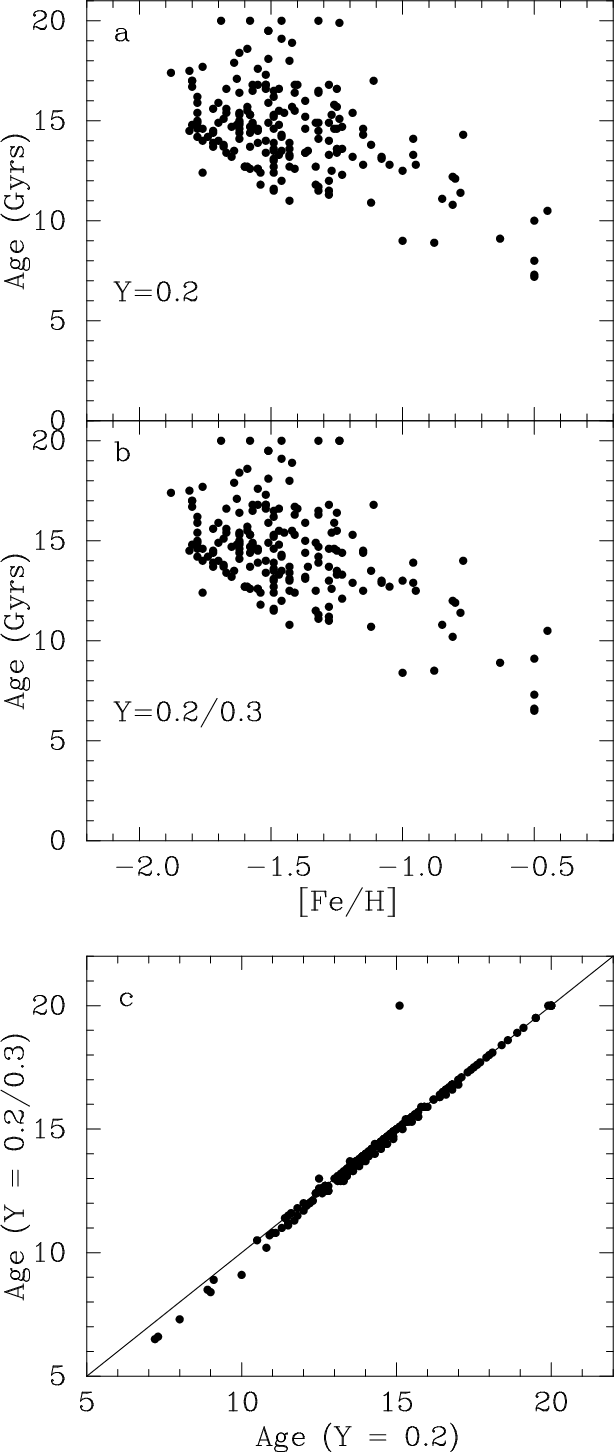}
\figcaption[f15.ps]{Age-metallicity relations resulting from ages
obtained using constant He abundance (Y=0.2) in panel~a, and varied
helium abundance (Y=0.2 for [Fe/H]$\leq$--1.5 and 0.3 for
[Fe/H]$>$--1.5) in panel~b for $V<$18.  Panel~c shows the comparison
between the ages calculated using the two different sets of
isochrones. See text (\S\ref{he_sect}) for details.
\label{ryi} }
\end{center}
\end{figure}

The results are shown in Figure~\ref{ryi}.  The two upper panels show
the AMDs for $\Delta$Y=0.0 and 0.1 respectively.  The lower panel
compares the ages determined with and without helium variations.
These plots show that there is little variation at the turnoff region
for a particular star when changing its assumed helium abundance.
Although there is a significant difference between the positions on
the CMD at the MS and RGB, the turnoff region difference is very
small.  We conclude that the possible enhancement of helium in the
second population does not significantly affect the ages calculated at
the turnoff region.  There is one star in which the ages do differ by
a significant amount (age = 15~Gyrs for Y=0.2 and age=20~Gyrs for
Y=0.3).  This is due to its position on the CMD.  It is at the SGB as
opposed to the turnoff region, and here the isochrones lie very close
together and small variations in position of the isochrones for the
two sets induce a large age variation.

\subsection{Method 2: Synthetic Color Magnitude Diagrams} \label{meth2_sect}

The second method to determine the age range in $\omega$ Cen utilized
only the photometry information.  No use was made of the spectroscopy
data other than it supports the use of \citetalias{nfm96} as the
abundance distribution.  Synthetic color-magnitude diagrams were
constructed using the Y$^{2}$ isochrones and the metallicity
distribution from \citetalias{nfm96}.  Monte Carlo simulations were
performed to obtain simulations with linear age ranges between 0 and
8~Gyrs, in 0.5 Gyr increments.  The age ranges were applied between
the metallicities [Fe/H]=--1.7 and~--0.6.  Any points that had
metallicities outside this range were assigned the maximum or minimum
value, with the maximum age being 13.5~Gyrs in each simulation. Errors
in photometry were assigned to each point, in accordance with the data
in Table~\ref{tbl-1}.  Each simulation had $N$=80000.  These
simulations did not include binary stars.

\begin{figure}
\begin{center}
\includegraphics[width=6.2cm,angle=0]{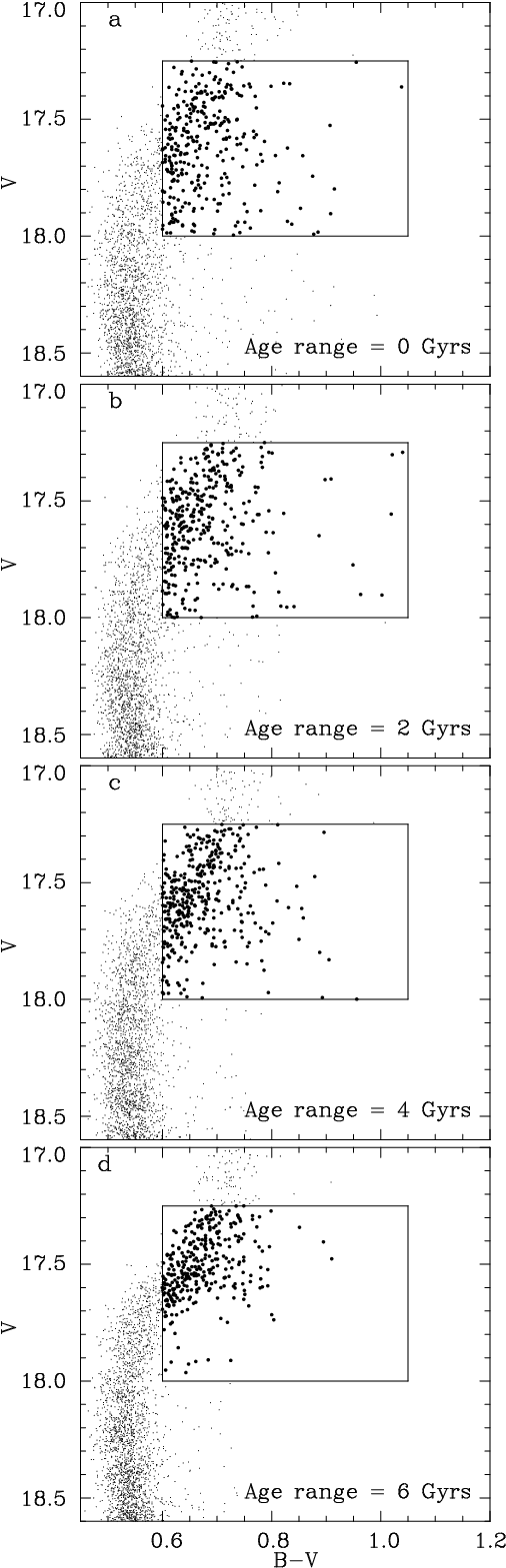}
\figcaption[f16.ps]{Examples of synthetic CMDs obtained
from four simulations. See text (\S\ref{meth2_sect}) for details.
\label{syncmd} }
\end{center}
\end{figure}

Four of the CMDs obtained from these simulations are shown in
Figure~\ref{syncmd}.  Each CMD shows the synthetic points, with the
objects with $V\leq$18 falling in the 2002 turnoff box as larger
symbols.  Figure~\ref{cmd3} shows the {\wcen} data and is for
comparison purposes.  Panel~a is for a simulation with no age spread;
b, c, and d are for age spreads of 2, 4 and 6~Gyrs, respectively,
between metallicities [Fe/H]=--1.7 and~--0.6.  In order to objectively
test which simulation represented our observed sample best, a
$\chi^{2}$ test was performed.  This involved dividing the CMD where
our objects lie into smaller boxes.  A grid of 3x5 boxes (three in
$V$~magnitude and five in $B-V$) was used in this case, and each box
was given equal weighting.  We established, through a series of tests
using different numbers of boxes, that the 3x5 grid gave the best
chance of finding the age range.

In order to accurately interpret the results of the $\chi^{2}$ fitting
we first tested our simulations and statistical calculations.  The
first test involved using a sample from one of the simulations to
represent our observed data.  The objects were chosen randomly and had
the same sample size as our observed sample ($n$=222).  This was
repeated five times to check the consistency of the results.  Four
representative samples were chosen with ages spreads of 0, 2, 4 and
6~Gyrs.  These were tested against all the other, larger N,
simulations in the same manner as the observed sample to see which
simulation was the ``best fit''.  The best result in these cases would
of course be the input age range.  For example, the 2~Gyr
representative sample should be best reproduced by the 2~Gyr
simulation.  The results of these tests are shown in
Figure~\ref{syntest}.  For the representative sample with no age
range, panel~a in Figure~\ref{syntest}, the statistical test seems to
find the correct result, although an age range of 2~Gyrs can not be
definitively ruled out.  The 2~Gyr representative sample (panel~b)
does not have a clear solution as to which simulation is best
represented by it and the test shows that it could be anywhere between
0 and 4~Gyrs.  The test does reproduce an age range of 4~Gyrs for the
4~Gyr representative sample (panel~c), but again there is uncertainty
of $\pm$2~Gyrs.  The representative sample with a 6~Gyr age spread
seems to be the most clearly defined result and accurately predicts an
age range of 6~Gyrs for the sample (panel~d).  These tests indicate
that the higher the age range the more likely it will be recovered by
this statistical test.  This effect was also seen in the simulations
done in \S\ref{meth1_sect}, and shown in Figure \ref{agemod}.  Lower
age ranges are harder to identify than larger ones.

The $\chi^{2}$ results for {\wcen} are shown in Figure~\ref{wcentest}.
Comparing these to the tests performed on the representative samples,
we can see that the {\wcen} results are similar to the 4~Gyr results.
Both have an age range of 4~Gyrs as the lowest $\chi^{2}$ value, and
both have small values for the 6~Gyr simulation as well.  The
$\chi^{2}$ result for the 2~Gyr age range is somewhat higher than in
the 4~Gyr graph, and resembles the 6~Gyr representative sample results
with a steep slope at low age ranges. The shape of the curve indicates
that the age range is not 0 or 2~Gyrs, but could possibly be up to
6~Gyrs. We can rule out large age ranges (of the order 6~Gyrs) as the
{\wcen} data do not show a distinct result for it in these
simulations, nor in \S\ref{meth1_sect}.

\subsection{Simulations with no age-metallicity relation} \label{noamr_sect}

\begin{figure}
\begin{center}
\includegraphics[width=6.2cm,angle=0]{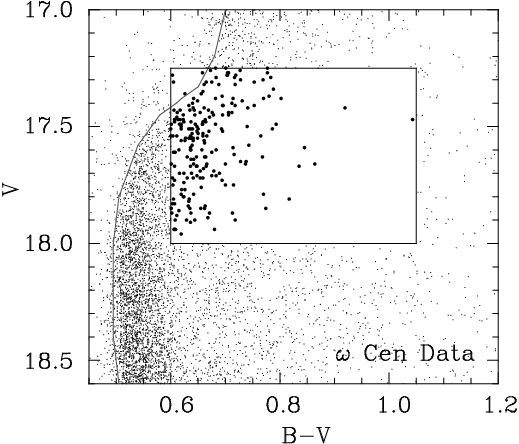}
\figcaption[f17.ps]{CMD for {\wcen} for comparison with the
synthetic simulations shown in Figure~\ref{syncmd}. The data plotted
here are only for stars in our sample that have $V<$18.  The solid
line indicates the blue fiducial edge of the {\wcen} data shown here,
relevant to the discussion in \S\ref{noamr_sect} and the accompanying
Figure~\ref{noage}.
\label{cmd3} }
\end{center}
\end{figure}

\begin{figure}
\begin{center}
\includegraphics[width=6.2cm,angle=0]{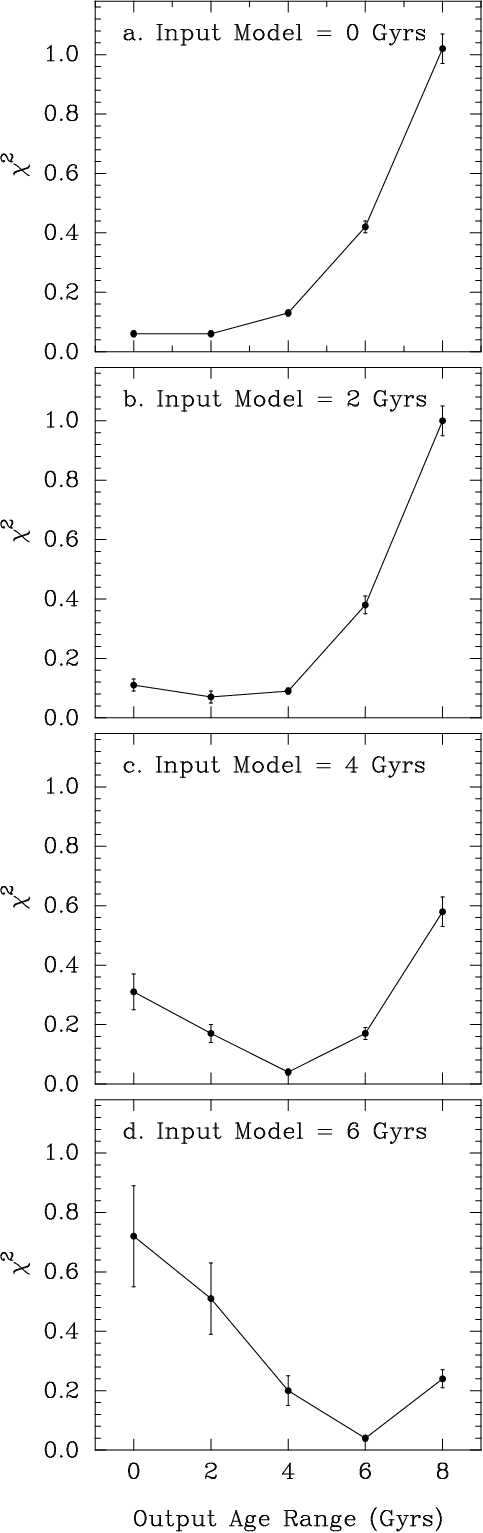}
\figcaption[f18.ps]{$\chi^{2}$ results for a set of
representative samples with age ranges of 0, 2, 4 and 6~Gyrs tested
against simulations of different age ranges. See text (\S \ref{meth2_sect})
for details.
\label{syntest} }
\end{center}
\end{figure}

\begin{figure}
\begin{center}
\includegraphics[width=6.2cm,angle=0]{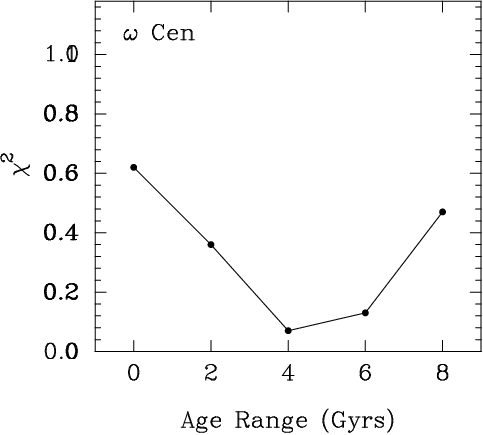}
\figcaption[f19.ps]{$\chi^{2}$ results for $\omega$
Cen.  A comparison between this and Figure~\ref{syntest} shows that the
age range in {\wcen} is most likely to be 4~Gyrs. \label{wcentest} }
\end{center}
\end{figure}

\begin{figure}
\begin{center}
\includegraphics[width=6.2cm,angle=0]{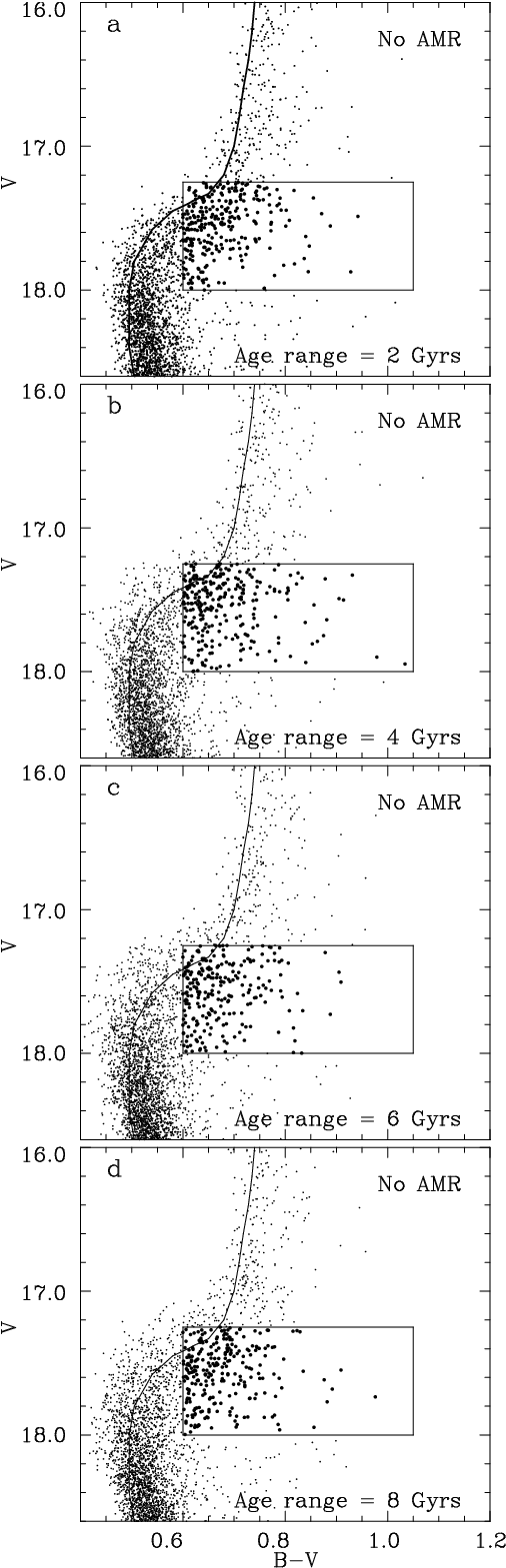}
\figcaption[f20.ps]{Examples of synthetic CMDs obtained
from four simulations.  Although the simulations have an age range,
they do not have an age-metallicity relation.  This means there can be
old metal-rich stars as well as young metal-poor stars in the
simulation. See text (\S~\ref{noamr_sect}) for details.
\label{noage} }
\end{center}
\end{figure}

Simulations were also produced that had the {\wcen} metallicity
distribution and a range in ages but no age-metallicity relation,
examples of which are shown in Figure~\ref{noage}.  The ages are
chosen randomly for each object, and as such had a flat distribution.
This figure shows four simulations with age ranges of 0, 2, 4 and
6~Gyrs.  The 2002 CMD limits are also shown, and the simulated objects
falling in this area are highlighted.  The solid line in the figures
represents the blue edge fiducial of the {\wcen} data (see
Figure~\ref{cmd3}).

These simulations did not match the 2002 sample.  In particular the
old metal-rich stars and young metal-poor stars were found in the
simulations but are not seen in the observational data. The young
metal-poor stars in the simulations are seen in the bluer and brighter
turnoff region which have no counterpart in the observed sample as
shown in Figure~\ref{cmd3}. The old metal-rich stars are found in the
bottom right corner of the 2002 box in the CMD and again there were
none of these stars found in the 2002 sample. This last result cannot
be explained by a lack of candidate members being observed in this
area. The simulated stars were well distributed across this region,
but no members of {\wcen} were found in that area. Although some of
these simulations might be considered as fitting the observed data, it
is the presence of young metal-poor stars at and above the turnoff
that excludes this scenario. This does indicate that there is a clear
age-metallicity relation in the cluster with the younger stars being
more metal-rich.

\section{Discussion} \label{disc_sect}

\subsection{Age Range}

We have shown using several different methods that there is a
difference in ages within the stars of {\wcen} of between 2 and
4~Gyrs.  Our first method of assigning individual ages to stars
initially seemed to indicate quite a large age range of 5~Gyrs.  It
was found, however, through simulations of the populations in {\wcen}
that this large age range was most probably induced by the
observational errors, which led to an enhanced age-metallicity
correlation.  Our simulations indicated that the actual age range in
the cluster was 2$\pm$2~Gyrs.  Simulations of the CMDs, using on the
photometry information, showed that the age range is 4$\pm2$~Gyrs.

Although our results do not give a definitive value for the age range
in {\wcen}, they can be used, along with other data, to strongly
constrain it.  Previous results for the existence of an age range in
$\omega$ Cen, and the method employed, are summarized in
Table~\ref{tbl-5}.  Considering these studies, an age range of
$\ge$6~Gyrs is most likely to be too high.  On the other hand,
although a zero age range can not be ruled out completely, it seems
unlikely to be the case, particularly given the results for the
element abundance ratios in the metal-rich populations. We therefore
conclude that the most likely value for the age spread in {\wcen} is
2--4~Gyrs.

Two results in the literature are most relevant to this work. The
first is that of \citet{hil04}.  They used Str\"{o}mgren photometry
and metallicities to determine an age range of 3~Gyrs.  Their result
is consistent to what we found here.

The second result is from \citet{sol05b} using two sets of photometry,
and metallicities derived from spectra.  They find little, if any, age
range (0--2~Gyrs) in their isochrone fits to the CMDs.  As we have
found an age range of 2--4 Gyrs these results are not entirely
inconsistent, but do differ enough to warrant further investigation.
Part of the explanation for this difference may come from the two
different regions of the cluster that we and \citet{sol05b} have
observed.  Our photometric data come from the outer regions of the
cluster between 15 and 25 arcminutes from the center, while the
\citet{sol05b} data originated from fields centered on the cluster out
to 10 arcminutes.  The metal-rich population is more centrally
concentrated \citepalias{nfm96}, and we may not be sampling enough of
these objects to make a conclusive statement on the ages of the most
metal-rich population.  In apparent disagreement with there being no
age range, however, is the lack of $\alpha$-enhancement in the most
metal-rich population \citep{pan02} suggesting supernovae Ia
involvement in the enrichment of the stars.  \citet{kob98} found the
progenitors of supernovae Ia take $<$1~Gyr to evolve, which is in
agreement with the result of \citet{sol05b}.  However, \cite{ytn96}
found that the lifetime of supernovae Ia progenitors is most likely to
be 0.5--3~Gyrs, which supports both the result found here and
\citet{sol05b}.  An abundance study of the s-process elements by
\citet{pan03} for the most metal-rich stars show enrichment on the
same scales as the metal-intermediate population ([s/Fe]$\sim$1.0 dex)
\citep{nd95, scl95, smi00}.  The sources of these enrichment processes
(AGB stars) take several gigayears to mature (\citealt{rom05}, and
references therein).

Examining Figures~3 and~5 from \citet{sol05b}, which are relevant to
their WFI data, one finds an age range of about 3 Gyrs is
possible. Using isochrones with slightly different metallicities than
those plotted by \citet{sol05b} (for example [Fe/H]=--1.85 for the MP
population and [Fe/H]=--1.2 for the MInt2 population, suggested by the
mean abundance in the sample range determined from their Figure~3),
ages of 17~Gyrs and 14~Gyrs are required to fit the MP and MInt2 data,
respectively, giving an age range of 3~Gyrs.  Further, the outliers in
the right panel of their Figure~5, which are explained by
\citet{sol05b} as possibly due to photometric or spectroscopic errors,
may instead require a much younger age.  It would be interesting to
know how far from cluster center these objects lie.  A possible
conclusion might be that there is a bimodality in age in the most
metal-rich population where the older metal-rich group resides
primarily in the center of the cluster, while the younger metal-rich
group resides in the outer regions.  Given the metallicity errors in
both \citet{sol05b} and our data sets, it is not possible to obtain a
definitive answer to this possibility.

In Figures~4 and~6 of \citet{sol05b}, which pertain to a different,
higher resolution data set, one does not find a similar age
difference, except to say there is a large range in metallicities
(shown in the upper panels of their Figure~4) for each population.
This might indicate that there is a range of ages in each population.
The lack of spread in the CMD for each metallicity group, however,
suggest that the single metallicity isochrone fits to these data are
an appropriate choice.

Throughout this work we have assumed the age-metallicity relation to
be a linear one. This may not be the case. We know that {\wcen} has at
least three (and possibly up to five) distinct populations.  The
length of time between the formation of these populations may not in
fact be linear. Unfortunately our data do not have the required
accuracy to address this question. More accurate metallicities are
required, for which higher resolution spectra are needed, as well as
larger samples of the most metal-rich population, to more accurately
determine the age range in the cluster.

\subsection{Cluster Origins}

The origin of {\wcen} is not well understood.  Due to its unique
metallicity and age ranges it is unlikely that it formed in the same
manner as other globular clusters. From the enrichment of its member
stars, it was massive enough to retain ejecta from AGB stars and
supernovae.  \citet{ts03} discuss the formation of globular clusters
as the result of cloud-cloud collisions. These collisions trigger star
formation, and chemical evolution in the resulting cluster depends on
the relative velocity of the initial clouds.  Those with low
velocities trigger star formation involving less than 1\% of the gas
and promote star formation episodes induced by supernovae, as would be
the case in {\wcen}.  On the other hand, collisions between clouds
with higher velocities do not retain enough gas to form later
generations after the initial star formation episode, resulting in
``normal'' globular clusters.

Alternatively, {\wcen} may be the result of mergers of several
globular clusters with discrete metallicities within the halo of the
Milky Way. However, several globular clusters of discrete metallicites
do not accurately reproduce the metallicity distribution seen on the
RGB (\citetalias{nfm96}, \citealt{smi00}), and the probability of
several clusters colliding and merging in the halo is low.  This
scenario is also not consistent with the s-process enhancements seen
in the more metal-rich populations.  A different twist to this
hypothesis is the {\it merger within a fragment} scenario
\citep{sea77,sz78}.  In the context of {\wcen}, the more metal-rich
component may have been another smaller cluster associated with the
parent dwarf galaxy that merged with the nucleus \citep{nor97,fer02}.

Current evidence suggests {\wcen} is most likely to be the remnant
nucleus of a dwarf spheroidal galaxy that was consumed by the Milky
Way \citep{fre93}, similar to the Sagittarius dwarf spheroidal galaxy
(dSph), which is currently in the process of being stripped.  The
similarities between {\wcen} and dSph galaxies was noted initially by
\citet{nb78}.  {\wcen} shows self-enrichment over a timescale of
several gigayears, consistent with this scenario.  The current orbit
of {\wcen} within the Galaxy is at odds with cluster enrichment over a
long timescale, suggesting that this is not where it initially formed.
In its current orbit, the Milky Way would not have allowed such
self-enrichment to take place since any gas would have been stripped
from the cluster not long after its ejection.  The retrograde motion
and small apocentric radius \citep{din99a} are unusual properties for
a globular cluster and further the suggestion that it did not
originally form in its current orbit. Frequent disk crossings would
also have stripped remaining gas from the cluster thereby stopping any
later star formation episodes.

\citet{bf03} have demonstrated using a self-consistent dynamical model
that {\wcen} could have been formed from a nucleated dwarf galaxy that
interacted and merged with the young Galactic disc over a period of
2.6~Gyrs. This model assumes that there is very little gas in the
Galactic disc at these times. The central nucleus survives tidal
stripping due to its compactness, and extended star formation is
induced by the Galactic tidal forces causing radial inflow which
triggers repetitive starbursts. Their Figure~4 shows that star
formation is enhanced slightly at several different epochs. This is
consistent with the age range found here for {\wcen} and the distinct
populations found in the cluster.

Numerical simulations have been used to analyze the dynamical
evolution of simulated dwarf galaxies that evolve to have the present
day kinematic characteristics of {\wcen} \citep{mcs03, chi04, mez05}.
As the disruption occurs, these systems may deposit large fractions of
their stars into the thick disc component of the Galaxy, leaving the
nucleus to orbit it.  As \citet{mcs03} note, the debris from the
progenitor may have already been found in the observations showing
signatures of merging events in the Milky Way \citep{gwn02}.  Analysis
of data from various surveys (such as the Sloan Digital Sky Survey
\citep{yor00} or RAdial Velocity Experiment \citep{ste03}) may find
the debris from the progenitor of {\wcen} in the thick disc, giving a
more comprehensive picture of the evolution of the cluster.

Abundance studies of the Sagittarius dwarf spheroidal galaxy have
shown similar patterns to those found in {\wcen} for some elements
\citep{msh05a, msh05b}, further supporting the idea that the cluster
was associated with an accreted dwarf galaxy.  Deficiencies in copper
are seen in both systems \citep{msh05a, cun02, pan02}, and abundance
patterns of other elements (La and Y) were also found to be
similar. [$\alpha$/Fe], Na and Al, however, exhibit different
patterns.  This last result indicates that although both systems share
a similar history, as one might expect, evolutionary differences exist
between them.  Care should be taken, however, when comparing {\wcen}
with dSph galaxies, as the relative sizes between the two systems are
quite different.  Despite this, comparisons can still be made between
the systems as {\wcen}, as the nucleus of a dSph, may have had gas
inflow from the parent dSph outer regions that was incorporated into
later generations of stars \citep{bn06}.

CMDs of other dwarf galaxies show episodic star formation episodes.
The Carina and Fornax dSphs show large ranges in both metallicity and
age (\citealt{mat98}, and references therein).  The Carina dwarf
spheroidal galaxy is known to have two, possibly three, episodes of
star formation \citep{sh94, hkmn98, mon03}. These results are
qualitatively similar to those found for {\wcen} --- star formation
over relatively large timescales, once again supporting the idea that
this cluster is the nucleated remnant of a dSph galaxy.

Another property of {\wcen} is the dependence of kinematics on
abundance.  \citet{nor97} showed that the metal-rich component in the
cluster is centrally concentrated and has a lower velocity dispersion
than the metal-poor population.  \citet{sol05b} not only find the
metal-intermediate populations have a lower velocity dispersion than
the metal-poor component, in agreement with \citet{nor97}, but further
find the most metal-rich population has a higher velocity dispersion
than the metal-intermediate, but not as large as the metal-poor.  The
Sculptor dSph galaxy shows the similar characteristic \citep{tol04} of
dependence of kinematics on abundance.  This dSph has two distinct
populations, one metal-rich ([Fe/H]~=~--1.4) and the other metal-poor
([Fe/H]~=~--2.0).  The higher metallicity stars show lower velocity
dispersion than the metal-poor component and are also more centrally
concentrated, just as reported for $\omega$ Cen.  \citet{tol04},
however, found no evidence of different systemic rotation between the
two components.  In contrast, in {\wcen} the metal-poor component
exhibits systemic rotation while the metal-rich one does not
\citep{nor97}, showing that there are differences between the systems
that are yet to be explained.  

\section{Conclusions}

Interpretation of our age-metallicity diagram is complicated by
correlated errors in metallicity and age, but after extensive
simulations an age range of 2--4~Gyrs is found to exist in {\wcen}. We
find an age-metallicity relation where the younger stars are those
that are more metal-rich.  These results strengthen the likelihood
that the origin of the unusual properties of this cluster is connected
with the evolution of a more massive system such as a nucleated dwarf
galaxy that was subsequently captured and disrupted by the Milky Way.

We thank the Director and staff of the Anglo Australian Observatory
for the use of their facilities.

\clearpage
\begin{table}
\begin{center}
\caption{Photometry errors for the {\wcen} data.
\label{tbl-1}}
\begin{tabular}{ccc}
\tableline\tableline
$V$ mag &  err$V$ & err($B-V$) \\
\tableline
16.0   & 0.002  &  0.004 \\ 
16.5   & 0.002  &  0.005 \\ 
17.0   & 0.004  &  0.007 \\ 
17.5   & 0.005  &  0.009 \\ 
18.0   & 0.008  &  0.012 \\ 
18.5   & 0.011  &  0.016 \\ 
19.0   & 0.017  &  0.025 \\ 
19.5   & 0.024  &  0.034 \\ 
20.0   & 0.031  &  0.045 \\ 
\tableline
\tableline
\end{tabular}
\end{center}
\end{table}

\clearpage
\begin{deluxetable}{ccccccccccc}
\tabletypesize{\footnotesize}
\tablecolumns{11}
\tablewidth{470pt}
\tablenum{2}
\tablecaption{$\omega$ Cen Members.  The full version of this table is available electronically. \label{tbl-2}}
\tablehead{\colhead{ID}&\colhead{RA}&\colhead{Dec}
&\colhead{Vmag}&\colhead{B--V}&\colhead{Vel$_h$}&\colhead{[Fe/H]}&\colhead{$\sigma_{[Fe/H]}$}&\colhead{Age}&\colhead{$\sigma_{Age}$}&\colhead{Run\tablenotemark{1}}\\
 & \scriptsize{J2000} & \scriptsize{J2000} & & & \scriptsize{kms$^{-1}$} & & & \scriptsize{(Gyrs)} & \scriptsize{(Gyrs)} &
}
\startdata
1000258 &  13 25 49.30 & --47 15 35.1 &  18.49 &  0.52 &  200 & --1.76 & 0.19 & 12.8 & 2.7 &  1\\ 
1000812 &  13 25 47.00 & --47 17 23.0 &  17.47 &  0.61 &  227 & --1.64 & 0.19 & 12.3 & 1.3 &  2\\ 
1002064 &  13 25 41.80 & --47 18 39.6 &  18.29 &  0.57 &  197 & --1.41 & 0.17 & 14.7 & 2.1 &  1\\ 
1002884 &  13 25 38.00 & --47 19 46.6 &  17.49 &  0.60 &  206 & --1.76 & 0.19 & 12.8 & 1.2 &  2\\ 
1004374 &  13 25 31.00 & --47 19 24.9 &  17.50 &  0.62 &  212 & --1.72 & 0.19 & 13.0 & 1.1 &  2\\ 
1005088 &  13 25 27.60 & --47 13 33.8 &  17.39 &  0.73 &  224 & --1.58 & 0.22 & 17.5 & 3.9 &  2\\ 
1005184 &  13 25 26.70 & --47 19 50.2 &  17.28 &  0.72 &  225 & --1.93 & 0.30 & 19.0 & 2.1 &  2\\ 
1005758 &  13 25 23.70 & --47 17 26.6 &  17.38 &  0.80 &  211 & --1.37 & 0.28 & 19.0 & 3.2 &  2\\ 
1005996 &  13 25 21.80 & --47 25 48.1 &  17.32 &  0.72 &  223 & --1.49 & 0.24 & 12.4 & 4.3 &  2\\ 
1006065 &  13 25 22.30 & --47 12 09.1 &  17.51 &  0.65 &  226 & --1.62 & 0.20 & 13.0 & 1.6 &  2\\ 
\enddata
\tablenotetext{1}{Observed in 98/99 (1), 2002 (2)}
\end{deluxetable}

\clearpage
\begin{deluxetable}{cccccrr}
\tabletypesize{\footnotesize}
\tablecolumns{7}
\tablewidth{315pt}
\tablenum{3}
\tablecaption{Non-Members in the 2002 observing box. The full version of this table is available electronically.  \label{tbl-3}}
\tablehead{\colhead{ID}&\colhead{RA}&\colhead{Dec}
&\colhead{Vmag}&\colhead{B--V}&\colhead{Vel$_h$}&\colhead{Run\tablenotemark{1}}\\
& \scriptsize{J2000} & \scriptsize{J2000} & & & \scriptsize{km s$^{-1}$} & 
}
\startdata
1001938 &  13 28 47.01 & --47 34 48.40 & 17.960 & 0.745 & \phm{00}83.82 &   2\\ 
1002364 &  13 28 20.23 & --47 27 21.50 & 17.454 & 0.935 & \phm{00}--69.53 & 2\\ 
1004333 &  13 28 31.77 & --47 19 56.80 & 17.256 & 0.988 & \phm{00}--69.60 & 2\\ 
1006419 &  13 26 52.11 & --47 11 47.70 & 17.963 & 0.957 & \phm{00}19.28 &   2\\ 
1006806 &  13 26 56.74 & --47 05 01.60 & 17.773 & 0.725 & \phm{00}51.16 &   2\\ 
1006842 &  13 26 40.69 & --47 05 41.80 & 17.567 & 0.785 & \phm{000}2.03 &   2\\ 
1007176 &  13 26 02.32 & --47 07 48.90 & 18.112 & 0.979 & \phm{000}--5.43 & 2\\ 
1007243 &  13 25 26.06 & --47 13 03.70 & 18.387 & 0.725 & \phm{00}--36.33 & 2\\ 
1007513 &  13 24 59.11 & --47 19 47.20 & 17.850 & 0.931 & \phm{00}--15.36 & 2\\ 
1008138 &  13 24 53.74 & --47 26 40.90 & 17.625 & 0.828 & \phm{00}--66.73 & 2\\ 
\enddata
\tablenotetext{1}{Observed in 98/99 (1), 2002 (2)}
\end{deluxetable}

\clearpage
\begin{table}
\begin{center}
\tablenum{4}
\caption{Parameters and observing dates for the calibrating clusters.
\label{tbl-4}}
\begin{tabular}{lccccl}
\tableline\tableline
Cluster & E($B-V$) & (m--M)$_{V}$ & [Fe/H]  & Number & Observed \\
\tableline
NGC 6397  & 0.17  &  12.36 & --1.95  &  111     & 23 Jul 98 \\
M 55      & 0.10  &  13.87 & --1.81  &   70     & 23 Jul 98 \\
          &       &        &         &          & 04 Jul 00 \\
NGC 6752  & 0.05  &  13.13 & --1.56  &  114     & 22 Sep 98 \\
          &       &        &         &          & 16 May 99 \\
          &       &        &         &          & 17 May 99 \\
47 Tuc    & 0.04  &  12.37 & --0.76  &  147     & 22 Nov 00 \\
          &       &        &         &          & 23 Nov 00 \\
\tableline
\end{tabular}
\end{center}
\end{table}

\clearpage
\begin{table}
\begin{center}
\tablenum{5}
\caption{Age Ranges in the Literature
\label{tbl-5}}
\begin{tabular}{llr}
\tableline\tableline
Reference    & Method     & Age Range \\
\tableline
\citet{nd95}  & Spect. of RGB stars        &             \\
              & s-process enrichment       & $\geq$ 1 Gyr \\
\citet{hr00}  & Str\"{o}mgren photometry   & 3--6 Gyrs   \\
\citet{hw00}  & Str\"{o}mgren photometry   & $\geq$ 3 Gyrs \\
\citet{smi00} & Spect. of RGB stars        & 2--3 Gyrs   \\
\citet{lee02} & Photometry, Red clump HB   & 4 Gyrs      \\
\citet{pan02} & Spect. of RGB stars        &             \\
              & SNe I enrichment           & $\leq$ 1 Gyr \\
\citet{ori03} & Spect. of RGB stars        &             \\
              & SNe I enrichment           & $\sim$ 1 Gyr \\
\citet{fer04} & High res. multiband phot.  &             \\
              & Subgiant isochrone fitting & 0 Gyrs      \\
\citet{hil04} & Spect. \& photometry       &             \\
              & of MSTO stars              & 3 Gyrs      \\
\citet{rey04} & BV, Ca, Str\"{o}mgren phot.  &      \\
              & Population models of HB    & 4 Gyrs      \\
\citet{sol05a} & RGB bumps                 & $<$ 6 Gyrs \\
\citet{sol05b} & SGB metallicities and     &               \\
               & isochrone fitting         & 0--2 Gyrs \\
This Paper    & Spectroscopy, photometry   &             \\
              & \& simulations of MSTO stars & 2--4 Gyrs      \\
\tableline
\tableline
\end{tabular}
\end{center}
\end{table}

\clearpage

\clearpage

\clearpage

\clearpage

\clearpage

\clearpage

\clearpage

\clearpage

\clearpage

\clearpage

\clearpage

\clearpage

\clearpage

\clearpage

\clearpage

\clearpage

\clearpage

\clearpage

\clearpage

\clearpage


\begin{thebibliography}{}
\bibitem[Bedin~et al.(2004)]{bed04}
Bedin, L. R.,  Piotto, G., Anderson, J., Cassisi, S., King, I. R., Momany, Y., \& Carraro, G. 2004, \apj, 605, L125
\bibitem[Beers~et al.(1999)]{bee99}
Beers, T. C., Rossi, S., Norris, J. E., Ryan, S. G., \& Shefler, T. 1999, \aj, 117, 981
\bibitem[Bekki~\&~Freeman(2003)]{bf03}
Bekki, K., \& Freeman, K. C. 2003, \mnras, 346, 11
\bibitem[Bekki~\&~Norris(2006)]{bn06}
Bekki, K., \& Norris, J. E. 2006, \apj, 637, 109
\bibitem[Cannon~(1981)]{can81}
Cannon, R. D., 1981, \mnras, 195, 1
\bibitem[Cannon~\&~Stewart(1981)]{cs81}
Cannon, R. D., \& Stewart, N. J. 1981, \mnras, 195, 15
\bibitem[Cannon~\&~Stobie(1973a)]{cs73}
Cannon, R. D., \& Stobie, R. S. 1973a, \mnras, 162, 207
\bibitem[Carney~et~al.(1996)]{clla96}
Carney, B. W., Laird, J. B., Latham, D. V., \& Aguilar, L. A., 1996, \aj, 112, 668
\bibitem[Chiba~\&~Mizutani(2004)]{chi04}
Chiba, M., \& Mizutani, A. 2004, PASA, 21, 237
\bibitem[Cunha~et~al.(2002)]{cun02}
Cunha, K., Smith, V. V., Suntzeff, N. B., Norris, J. E., Da Costa, G. S., \& Plez, B. 2002, \aj, 124, 379
\bibitem[Dinescu~et~al.(1999a)]{din99a}
Dinescu, D I., van Altena, W F., Girard, T M., \& López, C E. 1999a, \aj, 117, 277
\bibitem[Ferraro~et~al.(2002)]{fer02}
Ferraro, F. R., Bellazzini, M., \& Pancino, E. 2002, \apj, 573, L95
\bibitem[Ferraro~et~al.(2004)]{fer04}
Ferraro, F. R., Sollima, A., Pancino, E., Bellazzini, M., Straniero, O., Origlia, L., \& Cool, A. M. 2004, \apj, 603, L81
\bibitem[Freeman(1993)]{fre93}
Freeman, K. C. 1993, in Smith, G. H., \& Brodie, J. P., eds, ASP Conf. Ser. Vol. 48, The Globular Cluster-Galaxy Connection. (Astron. Soc. Pac.: San Francisco), p608
\bibitem[Freeman~\&~Rodgers(1975)]{fr75}
Freeman, K. C., \& Rodgers, A., W.  1975, ApJ, 201, L71
\bibitem[Freyhammer~et~al.(2005)]{fre05}
Freyhammer, L. M., Monelli, M., Bono, G., Cunti, P., Ferraro, I., Calamida, A., Degl'Innocenti, S., Prada Moroni, P. G., Del Principe, M., Piersimoni, A., Iannicola, G., Stetson, P. B., Andersen, M. I., Buonanno, R., Corsi, C. E., Dall'Ora, M., Petersen, J. O., Pulone, L., Sterken, C., \& Storm, J. 2005, \apj, 623, 860
\bibitem[Gilmore,~Wyse~\&~Norris(2002)]{gwn02}
Gilmore, G., Wyse, R. F. G., \& Norris, J. E. 2002, \apj, 574, L39
\bibitem[Green,~Demarque~\& King(1987)]{gdk87}
Green, E. M., Demarque, P., \& King, C. R. 1987, The Revised Yale Isochrones \& Luminosity Functions (New Haven: Yale Univ. Obs.)
\bibitem[Harris(1996)]{har96}
Harris, W. E., 1996, \aj, 112, 1478
\bibitem[Hilker~et~al.(2004)]{hil04}
Hilker, M., Kayser, A., Richtler, T., \& Willemsen, P.  2004, \aap, 442, L9
\bibitem[Hilker~\&~Richtler(2000)]{hr00}
Hilker, M., \& Richtler, T.  2000, A\&A, 362, 895
\bibitem[Hughes~\&~Wallerstein(2000)]{hw00}
Hughes, J., \& Wallerstein, G.  2000, AJ, 119, 1225
\bibitem[Hurley-Keller,~Mateo~\&~Nemec(1998)]{hkmn98}
Hurley-Keller, D., Mateo, M., \& Nemec, J. 1998, \aj, 115, 1840
\bibitem[Kim~et~al.(2002)]{kim02}
Kim, Y., Demarque, P., Yi, S., \& Alexander, D.  2002, ApJS, 143, 499
\bibitem[Kobayashi~et~al.(1998)]{kob98}
Kobayashi, C, Tsugimoto, T., Nomoto, K., Hachisu, I., \& Kato, M. 1998, \apj, 503, L159
\bibitem[Lee~et~al.(1999)]{lee99}
Lee, Y.-W., Joo, J.-M., Sohn, Y.-J., Rey, S.-C., Lee, H.-C., Walker, A. R. 1999, Nature, 402, 55
\bibitem[Lee~et~al.(2005)]{lee05}
Lee, Y.-W., Joo, S.-J., Han, S.-I., Chung, C., Ree, C. H., Sohn, Y.-J., Kim, Y.-C., Yoon, S.-J., Yi, S. K. \& Demarque, P. 2005, \apj, 621, L57
\bibitem[Lee~et~al.(2002)]{lee02}
Lee, Y.-W., Rey, S.-C., Ree, C. H., Joo, J. M., Sohn, Y.-J., Yoon,  S.-J., \& Walker, A. 2002, eds, ASP Conf. Ser. Vol. 265, A Unique Window into Astrophysics. (Astron. Soc. Pac.: San Francisco), p. 305
\bibitem[Lewis~et~al.(2002)]{lew02}
Lewis et al., 2002, \mnras, 333, 279
\bibitem[Lloyd~Evans(1977)]{le77}
Lloyd Evans, T. 1997, \mnras, 181, 591
\bibitem[Lub(2001)]{lub02}
Lub, J., 2002, in van Leeuwen, F., Hughes, J. D., \& Piotto, G. 2002, eds, ASP Conf. Ser. Vol. 265, A Unique Window into Astrophysics. (Astron. Soc. Pac.: San Francisco), p. 95
\bibitem[Mateo(1998)]{mat98}
Mateo, M. 1998, \araa, 36, 435
\bibitem[McWilliam~\&~Smecker-Hane(2005a)]{msh05a}
McWilliam, A., \& Smecker-Hane, T. A. 2005a, \apj, 622, L29
\bibitem[McWilliam~\&~Smecker-Hane(2005b)]{msh05b}
McWilliam, A., \& Smecker-Hane, T. A. 2005b, eds, ASP Conf. Ser. Vol. 336, Cosmic Abundances as Records of Stellar Evolution and Nucleosynthesis in honor of David L. Lambert. (Astron. Soc. Pac.: San Francisco), p. 221
\bibitem[Merritt,~Meylan,~\& Mayor~(1997)]{mmm97}
Merritt, D., Meylan, G., \& Mayor, M. 1997, AJ, 114, 107
\bibitem[Meza~et~al.(2005)]{mez05}
Meza, A., Navarro, J. F., Abadi, M. G., \& Steinmetz, M. 2005, \mnras, 359, 93
\bibitem[Mizutani,~Chiba~\& Sakamoto(2003)]{mcs03}
Mizutani, A., Chiba, M., \& Sakamoto, T. 2003, \apj, 589, L89
\bibitem[Monelli~et~al.(2003)]{mon03}
Monelli, M., Pulone, L., Corsi, C. E., Casterllani, M., Bono, G., Waker, A. R., Brocato, E., Buonanno, R., Caputo, F., Castellani, V., Dall'Ora, M., Marconi, M., Nonino, M., Ripepi, V., \& Smith, H. A. 2003, \aj, 126, 218
\bibitem[Norris(2004)]{nor04}
Norris, J. E. 2004, \apj, 612, L25
\bibitem[Norris~\&~Bessell(1978)]{nb78}
Norris, J., \& Bessell, M. S. 1978, \apj, 225, L49
\bibitem[Norris~\&~Da Costa(1995)]{nd95}
Norris, J. E., \& Da Costa, G. S.  1995, ApJ, 447, 680
\bibitem[Norris~et~al.(1997)]{nor97}
Norris, J. E., Freeman, K. C., Mayor, M., \& Seitzer, P. 1997, \apj, 487, L187
\bibitem[Norris,~Freeman~\& Mighell(1996)]{nfm96} 
Norris, J. E., Freeman, K. C., \& Mighell, K. J.  1996, ApJ, 462, 241
\bibitem[Origlia~et~al.(2003)]{ori03}
Origlia, L., Ferraro, F. R., Bellazzini, M., \& Pancino, E. 2003, \apj, 591, 916
\bibitem[Pancino(2003)]{pan03}
Pancino, E. 2003, Multiple Stellar Populations in $\omega$ Centauri, PhD thesis, University of Bologna, Italy
\bibitem[Pancino~et~al.(2000)]{pan00}
Pancino, E., Ferraro, F. R., Bellazzini, M., Piotto, G., \& Zoccali, M.  2000, ApJ, 534, 83
\bibitem[Pancino~et~al.(2002)]{pan02} 
Pancino, E., Pasquini, L., Hill, V., Ferraro, F. R., Bellazzini, M. 2002, \apj, 568, L101
\bibitem[Piotto~et~al.(2004)]{pio04}
Piotto, G., Villanova, S., Bedin, L. R., Gratton, R., Cassisi, S., Momany, Y., Recio-Blanco, A., Lucatello, S., Anderson, J., King, I. R., Peitrinferni, A. \& Carraro, G. 2005, \apj, 621, 777
\bibitem[Rey~et~al.(2004)]{rey04}
Rey, S.-C., Lee, Y.-W., Ree, C. H., Joo, M.-J., \& Sohn, Y.-J. 2004, \aj, 127, 958
\bibitem[Romano~et~al.(2005)]{rom05}
Romano, D., Chiappini, C., Matteucci, F., \& Tosi, M. 2005, \aa, 430, 491
\bibitem[Searle(1977)]{sea77}
Searle, L. 1977, in Tinsley, B. M., \& Larson, R. B. 1977, eds. The Evolution of Galaxies and Stellar Populations. Yale Univ. Obs., New Haven, p.219
\bibitem[Searle~\&~Zinn(1978)]{sz78}
Searle, L. \& Zinn, R. 1978, \apj, 225, 357
\bibitem[Smecker-Hane~et~al.(1994)]{sh94}
Smecker-Hane, T. A., Stetson, P. B., Hesser, J. E., \& Lehnert, M. D. 1994, \aj, 108, 507
\bibitem[Smith~et~al.(1995)]{scl95}
Smith, V., Cunha, K., \& Lambert, D.  1995, AJ, 110, 2827
\bibitem[Smith~et~al.(2000)]{smi00}
Smith, V., Suntzeff, N., Cunha, K., Gallino, R., Busso, M., Lambert, D., \& Straniero, O.  2000, AJ, 119, 1239
\bibitem[Sollima~et~al.(2005a)]{sol05a}
Sollima, A., Ferraro, F. R., Pancino, E. \& Bellazzini, M. 2005, \mnras, 357, 265
\bibitem[Sollima~et~al.(2005b)]{sol05b}
Sollima, A., Pancino, E., Ferraro, F. R., Bellazzini, M., Straniero, O., Pasquini, L. 2005, \apj, 634, 332
\bibitem[Stanford~et~al(2004)]{sta04}
Stanford, L. M, Da Costa, G. S., Norris, J. E., Cannon, R. D. 2004, Mem. Soc. Astron. Italiana, 75, 290
\bibitem[Steinmetz(2003)]{ste03}
Steinmetz, M. 2003, in GAIA Spectroscopy: Science and Technology, ASP Conf. Proc., 298, ed. U.~Munari, 381 
\bibitem[Suntzeff~\&~Kraft(1996)]{sk96}
Suntzeff, N. B., \& Kraft, R., P. 1996, \aj, 111, 1913
\bibitem[Tolstoy~et~al.(2004)]{tol04}
Tolstoy, E., Irwin, M. J., Helmi, A., Battaglia, G., Jablonka, P., Hill, V., Venn, K. A., Shetrone, M. D., Letarte, B., Cole, A. A., Primas, F., Francoise, P., Arimoto, N., Sadakane, K., Kaufer, A., Szeifert, T., \& Abel, T. 2004, \apj, 617, L119
\bibitem[Tsujimoto~\&~Shigeyama (2003)]{ts03}
Tsujimoto, T., \& Shigeyama, T. 2003, \apj, 2003, 590, 803
\bibitem[Woolley~et~al.(1966)]{woo66} 
Woolley, R. v. d. R., et al.\  1996, R. Obs. Ann., No 2
\bibitem[Yi~et~al.(2001)]{yi01}
Yi, S., Demarque, P., Kim, Y.-C., Lee, Y.-W., Ree, C.-H., Lejeune, T., \& Barnes, S. 2001, \apj, 136, 417
\bibitem[York~et~al.(2000)]{yor00}
York, D. G. et al. 2000, \aj, 120, 1579
\bibitem[Yoshii,~Tsujimoto,~\&~Nomoto(1996)]{ytn96}
Yoshii, Y., Tsujimoto, T., \& Nomoto, K. 1996, \apj, 462, 266
\end{thebibliography}
\end{document}